\documentclass[%
 reprint,
superscriptaddress,
nofootinbib,
 amsmath,amssymb,
 aps,
floatfix,
]{revtex4-1}

\usepackage{graphicx}
\usepackage{dcolumn}
\usepackage{bm}
\usepackage{hyperref}

\usepackage{xcolor}
\usepackage[normalem]{ulem}

\begin{document}

\title{Transport spectroscopy of ultraclean tunable band gaps in bilayer graphene}

\author{E.~Icking}
\affiliation{JARA-FIT and 2nd Institute of Physics, RWTH Aachen University, 52074 Aachen, Germany,~EU}%
\affiliation{Peter Gr\"unberg Institute  (PGI-9), Forschungszentrum J\"ulich, 52425 J\"ulich,~Germany,~EU}
\author{L.~Banszerus}%
\affiliation{JARA-FIT and 2nd Institute of Physics, RWTH Aachen University, 52074 Aachen, Germany,~EU}%
\affiliation{Peter Gr\"unberg Institute  (PGI-9), Forschungszentrum J\"ulich, 52425 J\"ulich,~Germany,~EU}%

\author{F. W\"ortche}
\affiliation{JARA-FIT and 2nd Institute of Physics, RWTH Aachen University, 52074 Aachen, Germany,~EU}%
\author{F.~Volmer}
\affiliation{JARA-FIT and 2nd Institute of Physics, RWTH Aachen University, 52074 Aachen, Germany,~EU}%

\author{P.~Schmidt}%
\affiliation{JARA-FIT and 2nd Institute of Physics, RWTH Aachen University, 52074 Aachen, Germany,~EU}%
\affiliation{Peter Gr\"unberg Institute  (PGI-9), Forschungszentrum J\"ulich, 52425 J\"ulich,~Germany,~EU}

\author{C.~Steiner}%
\affiliation{JARA-FIT and 2nd Institute of Physics, RWTH Aachen University, 52074 Aachen, Germany,~EU}%
\affiliation{Peter Gr\"unberg Institute  (PGI-9), Forschungszentrum J\"ulich, 52425 J\"ulich,~Germany,~EU}

\author{S.~Engels}%
\affiliation{JARA-FIT and 2nd Institute of Physics, RWTH Aachen University, 52074 Aachen, Germany,~EU}%
\affiliation{Peter Gr\"unberg Institute  (PGI-9), Forschungszentrum J\"ulich, 52425 J\"ulich,~Germany,~EU}

\author{J.~Hesselmann}%
\affiliation{JARA-FIT and 2nd Institute of Physics, RWTH Aachen University, 52074 Aachen, Germany,~EU}%

\author{M.~Goldsche}%
\affiliation{JARA-FIT and 2nd Institute of Physics, RWTH Aachen University, 52074 Aachen, Germany,~EU}%
\affiliation{Peter Gr\"unberg Institute  (PGI-9), Forschungszentrum J\"ulich, 52425 J\"ulich,~Germany,~EU}

\author{K.~Watanabe}%
\affiliation{
National Institute for Materials Science, 1-1 Namiki, Tsukuba, 305-0044, Japan }%

\author{T.~Taniguchi}%
\affiliation{
International Center for Materials Nanoarchitectonics, National Institute for Materials Science, 1-1 Namiki, Tsukuba 305-0044, Japan}%

\author{C.~Volk}
\affiliation{JARA-FIT and 2nd Institute of Physics, RWTH Aachen University, 52074 Aachen, Germany,~EU}%
\affiliation{Peter Gr\"unberg Institute  (PGI-9), Forschungszentrum J\"ulich, 52425 J\"ulich,~Germany,~EU}
\author{B.~Beschoten}
\affiliation{JARA-FIT and 2nd Institute of Physics, RWTH Aachen University, 52074 Aachen, Germany,~EU}%
\author{C.~Stampfer}
\affiliation{JARA-FIT and 2nd Institute of Physics, RWTH Aachen University, 52074 Aachen, Germany,~EU}%
\affiliation{Peter Gr\"unberg Institute  (PGI-9), Forschungszentrum J\"ulich, 52425 J\"ulich,~Germany,~EU}%

\date{\today}


\begin{abstract}

The importance of controlling both the charge carrier density and the band gap of a semiconductor cannot be overstated, as it opens the doors to a wide range of applications, including, e.g., highly-tunable transistors, photodetectors, and lasers. Bernal-stacked bilayer graphene is a unique van-der-Waals material that allows tuning the band gap by an out-of-plane electric field.
Although the first evidence of the tunable gap
was already found ten years ago, it took until recent to fabricate sufficiently clean heterostructures
where the electrically induced gap could be
used to fully suppress transport or confine charge carriers.
Here, we present a detailed study of the tunable band gap in gated bilayer graphene characterized by temperature-activated transport and finite-bias spectroscopy measurements. The latter method allows comparing different gate materials and device technologies, which directly affects the disorder potential in bilayer graphene. We show that graphite-gated bilayer graphene exhibits extremely low disorder and as good as no subgap states resulting in ultraclean tunable band gaps up to 120 meV. The size of the band gaps are in good agreement with theory and allow complete current suppression making a wide range of semiconductor applications possible.

\end{abstract}

\pacs{Valid PACS appear here}
\maketitle

\section{Introduction}

Bernal stacked bilayer graphene (BLG) is a unique material: intrinsically it is a 2D semi-metal, but it can be turned into a 2D semiconductor by applying an external out-of-plane electric field~\cite{McCann2006Mar,Min2007,McCann_2013}, with an electronic band gap that is directly related to the strength of the displacement field. 
The underlying mechanism of the band gap opening is a textbook example of how the breaking of inversion symmetry results in a gap in the electronic band structure.
The first experimental evidence of the tunable band gap in BLG was obtained by angle-resolved photoemission spectroscopy~\cite{Ohta2006,Zhou2007Oct} and infrared spectroscopy experiments~\cite{Mak2009, Kuzmenko2009Mar, Zhang2009Jun}, where band gaps up to 250~meV have been reported.
Signatures of the tunable band gap have also been observed
by scanning tunneling spectroscopy~\cite{Yankowitz2014Sep,Holdman2019Oct} and in early transport measurements~\cite{Oos2007Dec,Szafranek2010Mar,Xia2010Feb,Szafranek2011Jul,Lee2012Dec,Miyazaki2010,Yan2010Nov,Zou2010Aug,Taychatanapat_PhysRevLett2010,Jing2010Oct,Kanayama2015}. However, in the latter experiments, subgap states caused by disorder made it impossible to completely suppress the electron conduction~\cite{Connolly2012Nov}, making such BLG devices not suitable for semiconductor applications.

This shortcoming was solved neither by fabricating double-gated structures based on suspended BLG~\cite{Weitz2010,Allen2012Jul,Velasco2014, Ki2014} nor by encapsulating BLG into hexagonal boron nitride (hBN)~\cite{Zhu2017Feb,Goossens2012Aug,Shimazaki2015,Sui_nature2015}.
Only recently, with the use of graphite gates, the fabrication technology has advanced to the level where it is possible to open a gate-controlled band gap that results in a true band insulating state in BLG~\cite{Li2016,Overweg2018Jan}.

Here, we exploit this fabrication technology and show that the
tunable band gap of BLG can finally also be directly observed in finite bias transport spectroscopy measurements.
The obtained band gaps are in good agreement with theory as well as with the values extracted from thermally activated transport
~\cite{Shimazaki2015,Sui_nature2015,Taychatanapat_PhysRevLett2010,Zou2010Aug,Yan2010Nov,Jing2010Oct,Miyazaki2010,Kanayama2015}.
Most interestingly, we use
finite bias spectroscopy, to systematically compare different double-gated BLG/hBN device technologies, as this method allows to sensitively probe hopping-transport due to potential disorder or impurity states, which both can result in effective subgap and tail states.
The investigated devices differ mainly in the bottom gate material (graphite, gold, or highly doped silicon) and in the corresponding fabrication process. We show that the fabrication technology sensitively impacts the maximum device resistance and the presence and outline of diamonds of strongly suppressed conductance for finite bias voltage when measuring transport through electrostatically gapped BLG, as well as the tunability of the band gap with the electric displacement field.

We find that BLG devices with a graphite gate behave very closely to what theory predicts for ideal BLG, showing a truly semiconducting behavior in the presence of an applied displacement field. In the gapped regime, we measure maximum resistance values on the order of 100 G$\Omega$ (limited only by the measurement setup), and we observe no appreciable signature of trap or impurity states with subgap energy.
In contrast, both silicon and gold-gated devices appear to be affected by subgap states and disorder, but to different degrees. While very high gap-induced resistances are still observed in gold-gated devices, where the band gap just appears to be reduced in finite bias measurements, no gap can be directly observed in silicon-gated devices.
All this confirms that the intrinsic properties of BLG become exploitable in graphite-gated BLG/hBN heterostructures, which therefore represents the most promising platform to unleash the potential of this unique tunable 2D semiconductor with 
interesting applications for THz electronics, quantum technologies, and mesoscopic physics.

\begin{figure}[t!]
	\includegraphics[width=1.0\linewidth]{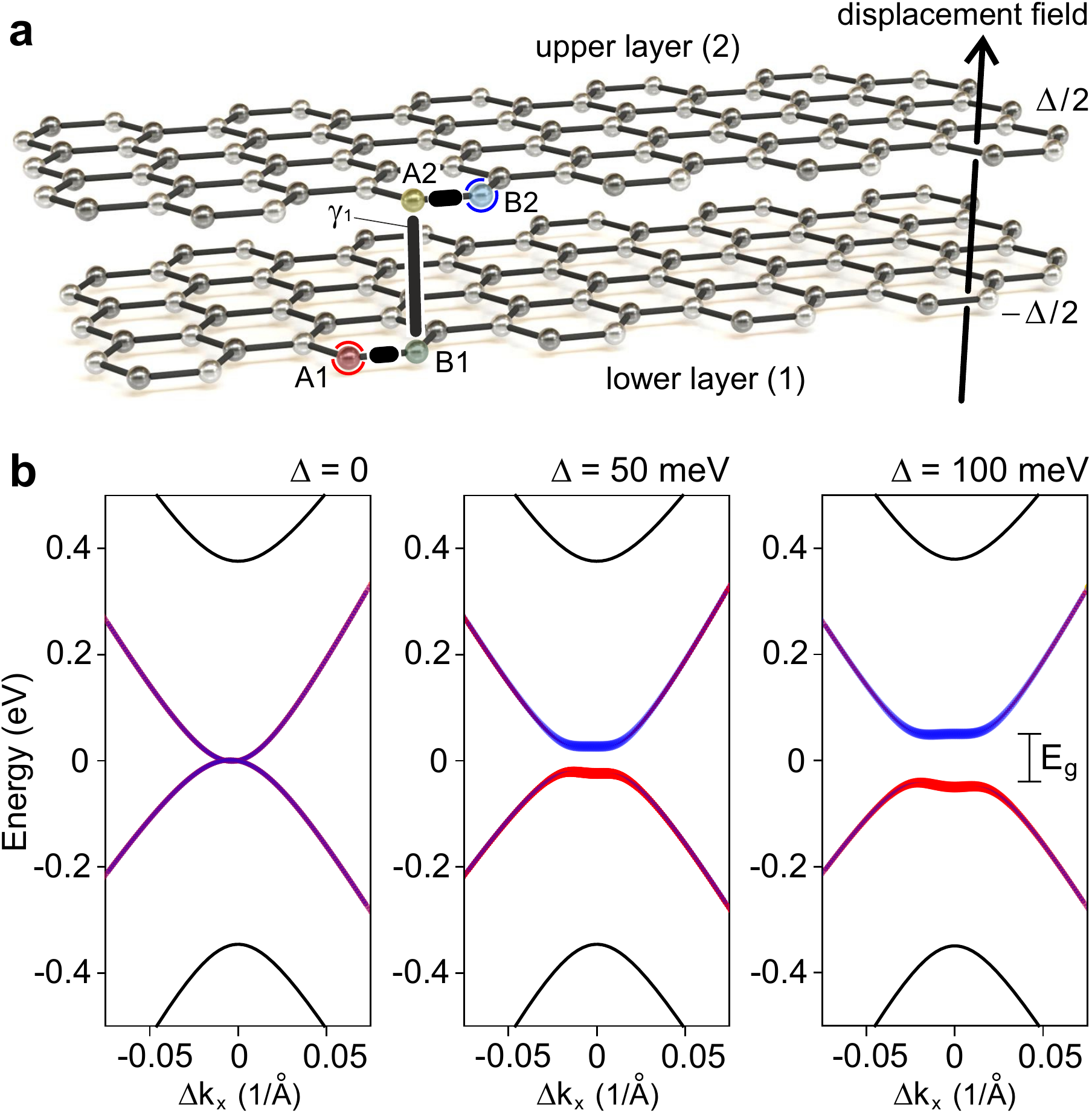}
	\caption{\textbf{(a)} Schematic illustration of Bernal-stacked bilayer graphene. An out-of-plane electric displacement field results in the onsite potential difference $\Delta$ between the layers, which breaks the inversion symmetry. The atomic orbitals of atoms A1 and B2 (highlighted in red and blue) determine the low energy spectra.
	\textbf{(b)} Electronic band structure of BLG near the K-point, calculated according to Equation~(30) in Ref.~\cite{McCann_206} for zero on-site potential difference (left panel), $\Delta=50$~meV (central panel), and $\Delta=100$~meV (right panel). The projection on the $2p_z ^\mathrm{A1}$ and $2p_z ^\mathrm{B2}$-orbitals are highlighted in blue and red, respectively. 
	}
	\label{fig:exp_0}
\end{figure}

\begin{figure*}[ht]
	\includegraphics[width=1.0\linewidth]{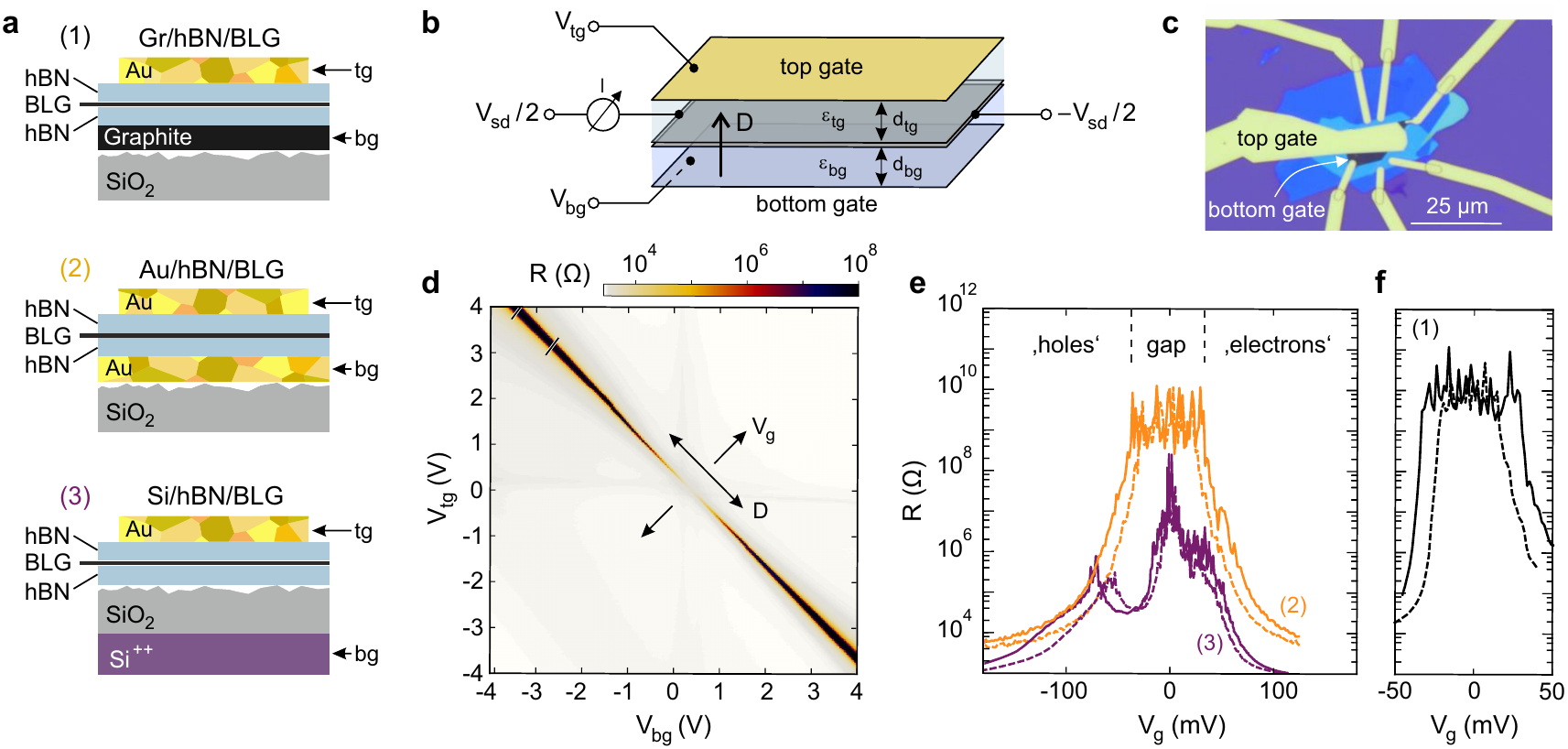}
	\caption{\textbf{(a)}
	Schematic cross-sections of the three types of double-gated BLG devices (class 1-3) investigated in this work. All devices have a gold top gate (tg) but different bottom gates (bg).
	\textbf{(b)} Illustration of a generic double-gated BLG device consisting of a top gate and a bottom gate
	and of the contacting scheme.
    \textbf{(c)} Optical image of an Au/hBN/BLG device. \textbf{(d)} Two-terminal resistance of the device shown in panel (c) as a function of $V_\mathrm{bg}$ and $V_\mathrm{tg}$ at $T=1.6$~K and $V_\mathrm{sd}=1~$mV. \textbf{(e)} Resistance of the Au/hBN/BLG device (orange lines) and of the Si/hBN/BLG device (purple lines) measured at $T=1.6$~K and $V_\mathrm{sd}=2~$mV along the direction of the short black lines in panel (d). These lines correspond to displacement fields $D=-0.58$~V/nm and $D=-0.48~$V/nm for the Au/hBN/BLG device (solid and dashed orange line, respectively), and to $D=-0.56~$V/nm and $D=-0.48~$V/nm for the Si/hBN/BLG device (solid and dashed purple line, respectively). \textbf{(f)} Same type of measurement  as in panel (e) for a Gr/hBN/BLG device, for displacement fields $D=-0.54$~V/nm (solid line) and $D=-0.46$~V/nm (dashed line).
	}
	\label{fig:exp_01}
\end{figure*}

\begin{figure*}[t]
	\includegraphics[width=1\linewidth]{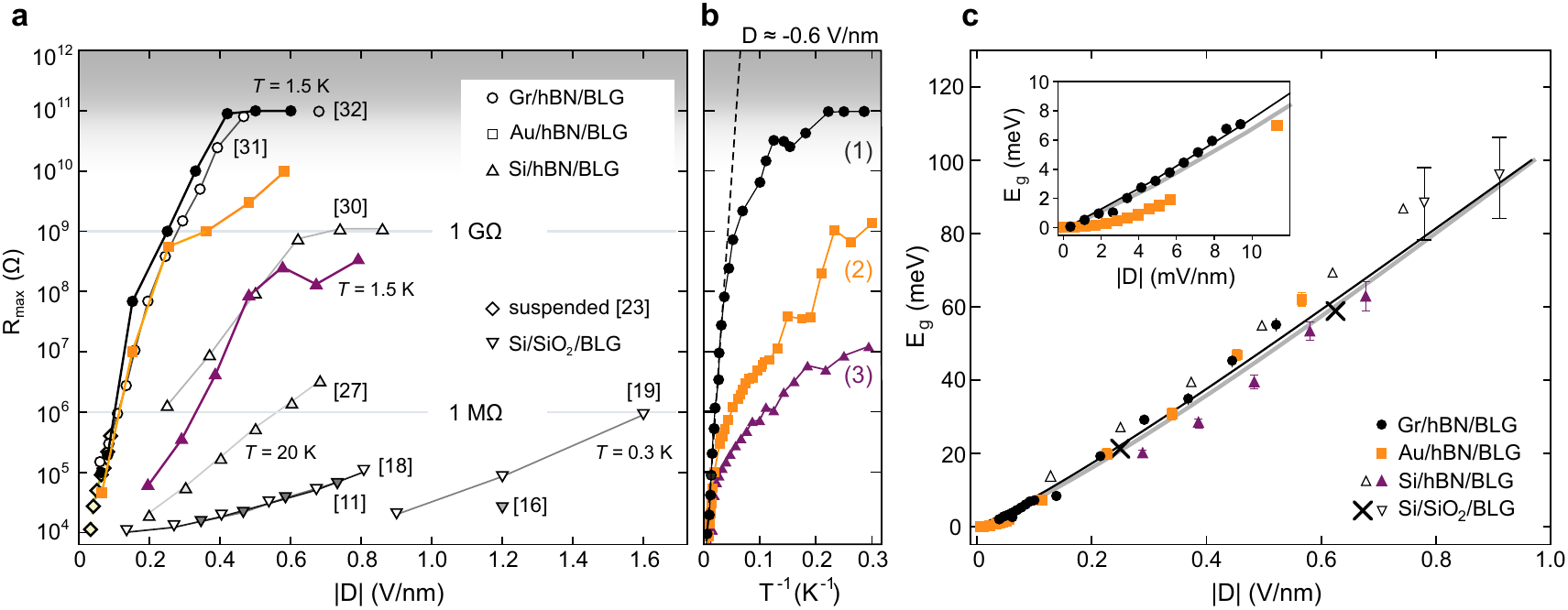}
	\caption{
	\textbf{(a)} Maximum resistance ($R_\mathrm{max}$) as a function of displacement-field for 12 different devices based on different gating technologies (see labels). This plot include both data from the devices investigated in Figure~2 (colored data points, same color code as in Figures~2e,f), as well as data from literature.
	The dark gray area shows resistance values that are not measurable with our experimental setup. Therefore only a lower limit can be given.
	\textbf{(b)} Arrhenius plot showing the maximum resistance $R_\mathrm{max}$ of the three devices investigated in Figures~2e,f as a function of the inverse temperature $1/T$ at a constant $D$-field (same labeling as in Figures~2e,f, and panel (a)).
    The black dashed line corresponds to a fit of the maximum resistance assuming thermal activated transport $R_\mathrm{max} \propto \exp(E_\mathrm{g}/(2 k_B T))$.
	\textbf{(c)} Band gap energy $E_\mathrm{g}$ as function of the displacement field. The data points correspond to the values of $E_\mathrm{g}$ extracted by fitting the high-temperature regime of the Arrhenius plot. The black line indicates the values of $E_\mathrm{g}$ predicted by the model of McCann and Koshino~\cite{McCann_2013}, the gray line shows the solution of a simplified model by Ref.~\cite{Slizovskiy2021} with $\varepsilon_\text{z}=1.65$ (see Supplementary Material).
	The black crosses are taken from Ref.~\cite{Zhang2009Jun}, the open downward-pointing triangles from Ref.~\cite{Yan2010Nov} and the open upward-pointing triangles from Ref.~\cite{Sui_nature2015}. The inset shows a close-up for small $D$-fields.
	}
	\label{fig:exp_02}
\end{figure*}

\section{An electrostatically tunable band gap}

The inversion symmetry and its controlled breaking play an important role in determining the properties of intrinsic and gapped BLG.
In the intrinsic form, the orbitals of the carbon atoms A1 and B2, which are responsible for the low energy spectra of BLG, are inversion-symmetric, and BLG is a semi-metal (see Figures~1a,b)~\cite{Min2007,McCann_2013}.
The symmetry is broken in the presence of an external out-of-plane displacement field, which induces an onsite potential difference $\Delta$ between the upper and lower graphene layer. This potential difference leads in turn to the appearance of a band gap between the conductance and the valence band~\cite{McCann2006Mar,Min2007,McCann_2013}, as illustrated in Figure~\ref{fig:exp_0}b.
The size of the band gap, $E_\mathrm{g}$, depends on the onsite potential difference $\Delta$ as~\cite{McCann_2013,Min2007}:
\begin{equation}
E_\mathrm{g}(\Delta)=\frac{|\Delta|}{\sqrt{1+(\Delta/\gamma_1)^2}},
\end{equation}
where $\gamma_1\approx 0.38$~eV is the interlayer coupling strength~\cite{Kuzmenko2009Nov,Joucken2020,Jung2014Jan,McCann_206}.
The dependence between $\Delta$ and the external electric displacement field is however non-trivial, as $\Delta$ depends also on the screening of the charge carriers on the layers of BLG, which is influenced in turn by the onsite potential difference $\Delta$, thus requiring a self-consistent analysis~\cite{McCann_2013,Slizovskiy2021}.

Different models have been used to calculate the dependence of the interlayer asymmetry $\Delta$ on the applied out-of-plane displacement field $D$, either using a simple
plate-capacitor model with Hartree screening~\cite{McCann2006Mar,McCann_2013}
or, more recently, by additionally taking into account the layer-dependent out-of-plane polarization of the carbon orbitals ~\cite{Slizovskiy2021}.
In both cases
$\Delta$ can be expressed as
\begin{equation}
\label{DeltaMT}
    \Delta=\frac{ d_0 e D}{\varepsilon_0\varepsilon_\text{z}}+\frac{d_0 e^2}{2\varepsilon_0 \overline{\varepsilon}}\delta n(\Delta),
\end{equation}
where $d_0=0.34$~nm is the interlayer spacing of BLG, $e$ is the (magnitude of the) elementary charge, and $\varepsilon_\text{z}, \overline{\varepsilon}$ are effective dielectric constants.
In the model of Ref.~\cite{McCann2006Mar,McCann_2013}, $\varepsilon_\text{z}=\overline{\varepsilon}=\varepsilon_\mathrm{BLG}$, where $\varepsilon_\mathrm{BLG} \approx 2$
is the effective dielectric constant of BLG~\cite{Kumar2016,Jung2017,Island2019}.
In contrast, in Ref.~\cite{Slizovskiy2021} $\varepsilon_\text{z}$ is the effective out-of-plane dielectric susceptibility of graphene and $\overline{\varepsilon}\equiv 2/(1+\varepsilon_\text{z}^{-1})$. Finally, $\delta n(\Delta)$ is the difference between the charge carrier density in  the upper and lower layer, whose detailed expression also depends on the considered model (see Supplementary Material). Below we will make use of both models when comparing theory with experiment.

\subsection{Double-gated BLG devices}
Experimentally,
the way to apply an out-of-plane electric field to BLG and to control independently its chemical potential is to embed it into a plate capacitor, i.e., to have a bottom and a top gate (see Figures~\ref{fig:exp_01}a,b).
In this work, we compare devices fabricated with three different technologies. All devices are based on BLG encapsulated into hexagonal boron nitride and have a metallic top gate (tg), but differ in the bottom gate (bg), as illustrated in Figure~\ref{fig:exp_01}a. Specifically, we consider devices with a graphite bottom gate (referred to as ``Gr/hBN/BLG" or class 1 devices), devices with a gold (Au) bottom gate (referred to as ``Au/hBN/BLG" or class 2 devices), and devices that use the heavily doped silicon substrate as bottom gate (referred to as ``Si/hBN/BLG" or class 3 devices). An optical image of a final device with an Au bottom gate
based on a stack-flipping process is shown in Figure~2c.
It is worth noting here that although the Au gate/stack-flipping process results in less disorder than directly placing an hBN/BLG/hBN stack on a (rough) Au bottom gate, the Au top gate (used in all devices) has overall less detrimental effect on device quality, which is due to the fact that the BLG heterostructure no longer needs to be moved afterwards.
Details on the device fabrication are given in the Supplementary Material.

In the investigated devices, the top gate is narrower than the bottom gate, and the BLG regions doped exclusively by the bottom gate act as leads.
In the double-gated region, the voltages $V_{\rm tg}$  and $V_{\rm bg}$ applied to the top and to the bottom gate induce a displacement field $D$ through the BLG:
\begin{equation}
    D=\frac{e}{2}\left[\alpha_\mathrm{bg}\left(V_\mathrm{bg}-V^0_\mathrm{bg}\right)-\alpha_\mathrm{tg}\left(V_\mathrm{tg}-V^0_\mathrm{tg}\right)\right],
\end{equation}
where
$V^0_\mathrm{tg}, V^0_\mathrm{bg}$ are the offset of the charge neutrality point (CNP) from $V_\mathrm{tg}=V_\mathrm{bg}=0$, and $\alpha_\mathrm{tg}=\varepsilon_0 \varepsilon_\mathrm{tg} /(e d_\mathrm{tg})$ and $\alpha_\mathrm{bg}=\varepsilon_0 \varepsilon_\mathrm{bg} /(e d_\mathrm{bg})$ are
the lever-arms determined by the capacitive coupling of the top and bottom gate, respectively ($\varepsilon_\mathrm{tg}$, $ \varepsilon_\mathrm{bg}$ are the dielectric constants and $d_\mathrm{tg}$, $d_\mathrm{bg}$ the thicknesses of the dielectric layers; see Figure~2b).
The effect of the voltages $V_{\rm tg}$ and $V_{\rm bg}$ on the chemical potential in the double-gated region is understood at best in terms of the effective gate voltage
\begin{equation}
\label{V_g}
V_\mathrm{g} = \frac{(V_\mathrm{bg} - V_\mathrm{bg}^\mathrm{0}) + \beta \; (V_\mathrm{tg} - V_\mathrm{tg}^\mathrm{0})}{1+\beta},
\end{equation}
where $\beta=\alpha_\mathrm{tg}/\alpha_\mathrm{bg}$ is the
ratio of the two lever-arms. This effective gate voltage is defined such that $V_\mathrm{g}$ is directly linked to the electro-chemical potential via $\mu \approx e\, V_\mathrm{g}$, as long as $\mu$ is within the band gap. Also $V_\mathrm{g}$ allows to change the total charge carrier density $n$ outside the band gap by $\Delta n \approx (\alpha_\mathrm{tg}+\alpha_\mathrm{bg}) \Delta V_\mathrm{g}$, if both layers are on the same potential (for more details see Supplementary Material).
For all devices, we extract the values of $V^0_\mathrm{tg}, V^0_\mathrm{bg}$ and $\beta$ from resistance-map measurements, and the value of either $\alpha_\mathrm{tg}$ or $\alpha_\mathrm{bg}$ from quantum Hall measurements~\cite{Zhao2010,SonntagMay2018,Schmitz2020} (see Supplementary Material).

\subsection{Resistance maps and maximal resistance}
As a first step to characterize the double-gated BLG devices, we record resistance maps like the one of Figure~\ref{fig:exp_01}d, by applying a small source-drain bias $V_\mathrm{sd}$ and by measuring the current $I$ as a function of $V_\mathrm{tg}$ and $V_\mathrm{bg}$ (see schematic in Figure~2b). The resistance maps exhibit two distinct features: (I)~A diagonal line of elevated resistance $R$, which marks the shifting of the charge neutrality point of the BLG in the double-gated region as a function of $V_\mathrm{tg}$ and $V_\mathrm{bg}$. This line represents the $V_\mathrm{g}=0$ axis (i.e., $\mu=0$). The slope of this line is directly given by the ratio of the lever arms $\beta=\alpha_\mathrm{tg}/\alpha_\mathrm{bg}$. For example, for the measurement shown in Figure~\ref{fig:exp_01}d, it is $\beta=0.99$, indicating a very symmetric capacitive coupling of the top and bottom gate.
(II)~Along this diagonal line the maximum resistance $R_\mathrm{max}$ strongly increases while increasing the magnitude of the displacement field, which is a hallmark of the opening of the band gap induced by the displacement field.

While these qualitative features are common to the resistance maps performed on devices of all three different classes (for a comparison of carrier mobilities see Supplementary Material), the differences between the three technologies become apparent when comparing line traces of the resistance measured as a function of $V_\mathrm{g}$ at fixed $D$-fields, see Figures~\ref{fig:exp_01}e,f.
In all cases, we observe an abrupt decrease of the largest resistance around $V_\mathrm{g}=0$. But, while devices with a graphite and gold bottom gate show well-defined plateaus of high resistance (see Figure~2f and orange traces in Figure~2e), the resistance of the device with a silicon bottom gate is significantly lower and varies by more than three orders of magnitude within the gate voltage range where the band gap is expected (purple traces in Figure~\ref{fig:exp_01}e). These strong variations and the reduced maximum-resistance can be explained in terms of hopping transport through subgap states caused by the disorder due to charged impurities in the SiO$_2$, at the SiO$_2$/hBN interfaces, or at the unscreened BLG edges~\cite{Oos2007Dec,Miyazaki2010,Zou2010Aug,Yan2010Nov,Taychatanapat_PhysRevLett2010,Kanayama2015,Zhu2017Feb} which create spatial electrostatic variations along the transport channel and are the main origin of the observed disorder potentials (see also discussion in Sec.~IV).
Vice versa, the high-resistance plateaus observed in devices with graphite or a gold bottom gate suggest that the chemical potential is tuned through a clean band gap, with few or no subgap states. For both types of devices, the width of the high-resistance plateau -- i.e., the size of the band gap -- increases as expected with increasing displacement field (see dashed and solid resistance traces in Figure~\ref{fig:exp_01}e,f).
However, devices with a graphite bottom gate exhibit a much sharper onset of the plateau and slightly higher values of resistance than those with a gold bottom gate, indicating that graphite gates are more effective than metallic ones at suppressing residual charge transport through the band gap and particularly near the band edges. This difference can be attributed to the disorder, i.e. spatial electrostatic variations along the BLG transport channel, caused (I) by the fabrication process of the Au-gated devices, which is more prone to interface contamination, or (II) ultimately by grain boundaries in the gold itself~\cite{Osvald1992}.

The different performance of the three technologies becomes even more apparent by plotting the maximum resistance $R_\mathrm{max}$ as a function of the displacement field $D$, see Figure~\ref{fig:exp_02}a (for more details on why $R_\mathrm{max}$ is a good quantity, see Supplementary Material). In this plot, we report both data from three devices measured in our lab (colored data points), as well as data taken from literature (open data points).
This overview plot shows that the maximum resistance attainable for a given value of $D$ strongly depends on the fabrication technology used for the double-gated BLG devices. Devices, where BLG is directly placed on SiO$_2$, require high displacement fields $\approx 1.6$~V/nm to reach  moderate values of $R_\mathrm{max} \approx$ 1~M$\Omega$.
Encapsulating BLG between hBN layers already helps reducing the disorder potential and allows achieving maximum resistances in the range of 100~M$\Omega$~$-$~1~G$\Omega$ at $D$-fields of around $|D|approx0.6~$ V/nm (and a temperature of $T \approx 1.5$~K).
The observed saturation of $R_\mathrm{max}$ for higher $D$-fields can be explained by disorder-induced hopping transport at subgap energies and along the edges of the BLG~\cite{Oos2007Dec,Miyazaki2010,Zou2010Aug,Yan2010Nov,Taychatanapat_PhysRevLett2010,Kanayama2015,Zhu2017Feb}.
The use of a metallic gate allows to reach values of $R_\mathrm{max}$ as high as 10~G$\Omega$ at moderate values of $D$, but it is only the use of a graphite gate that allows to open a real clean band gap and to completely suppress the current at reasonably low $D$-fields, reaching $R_\mathrm{max}\approx 100~$G$\Omega$ -- which represents also the maximum value of resistance measurable in our experimental setup. The values of $R_\mathrm{max}$ measured in our Gr/hBN/BLG device compare well with the data of Ref.~\cite{Overweg2018Jan,Li2016} and, at low $D$-fields, also with the values measured in a double-gated suspended BLG device (see diamond symbols in Figure~\ref{fig:exp_02}a, taken from Ref.~\cite{Weitz2010}).

\subsection{Thermally activated transport}

The maximum resistance is, of course, a function of temperature. For the three devices illustrated in Figure~2a, we study the maximum resistance $R_\mathrm{max}$ as a function of temperature for a fixed value of the displacement field $D$. To extract information on the underlying transport mechanisms, we plot $R_\mathrm{max}$ logarithmically as a function of $T^{-1}$, as shown in Figure~3b.

At low temperatures, $T<10$~K ( $T^{-1}>0.1$~K$^{-1}$), $R_\mathrm{max}$ only weakly depends on $1/T$, as predicted by both variable range hopping (VRH)~\cite{Yan2010Nov,Jing2010Oct} and a combination of nearest-neighbor hopping and VRH~\cite{Sui_nature2015,Zou2010Aug}.
In this regime, the values of $R_\mathrm{max}$ observed for the three gating technologies differ by several orders of magnitude.
The low resistance values at low temperature of the Si/hBN/BLG sample are reminiscent of those observed in disordered semiconductors~\cite{Shklovskii1984}, where transport via impurity bands and hopping transport dominates. 
This notion is also in agreement with earlier studies~\cite{Miyazaki2010,Kanayama2015} and with compressibility measurements~\cite{Henriksen2010Jul,Young2012Jun}, which have shown that there is a large density of (localized) states in gapped BLG when placed on SiO$_2$, resulting in low values of $R_\mathrm{max}$.

At high temperatures $T > 40$~K (T$^{-1} < 0.025$~K$^{-1}$), for all types of devices the dependence of $R_\mathrm{max}$ on $1/T$ is well described by thermally activated transport~\cite{Shimazaki2015,Sui_nature2015,Taychatanapat_PhysRevLett2010,Zou2010Aug,Yan2010Nov,Jing2010Oct,Miyazaki2010,Kanayama2015}, i.e., $R_\mathrm{max} \propto \exp(E_\mathrm{g}/(2 k_\mathrm{B} T))$, where $E_\mathrm{g}$ is the band gap energy and $k_\mathrm{B}$ the Boltzmann constant.
By fitting a line to the resistance data in the Arrhenius plot 
(see, e.g., dashed line in Figure~3b) we can extract $E_\mathrm{g}$ at a given value of $D$ for the different devices. Repeating this type of fitting for different values of $D$, we obtain the plot shown in Figure~3c. It can be observed, that the values of $E_\mathrm{g}$ determined in this way for the different devices agree rather well with each other, independently of the fabrication technology. This is the case because at high temperatures the impact of localized subgap states is eventually smeared out by thermal excitations.

Our data agree also well with the band gaps reported by earlier experiments (see opens symbols and crosses in  Figure~3c)~\cite{Zhang2009Jun,Yan2010Nov, Sui_nature2015} and with the values predicted by theory. Here, we consider both the self-consistent solution of the plate-capacitor model with Hartree screening proposed by McCann and Koshino~\cite{McCann_2013} with $\varepsilon_\mathrm{BLG}=2$ (black line in Figure~3c), as well as the more recent model of Ref.~\cite{Slizovskiy2021}, which takes into account the layer-dependent polarization of the orbitals (gray line). In this case, we used $\varepsilon_\text{z}=1.65$ for the effective out-of-plane dielectric susceptibility of graphene, see Equation~(2). For the device with graphite back gate, the extracted values of $E_\mathrm{g}$ agree well with theory even at very low displacement fields (see inset in Figure~3c).


\begin{figure*}[t]
	\includegraphics[width=0.95\linewidth]{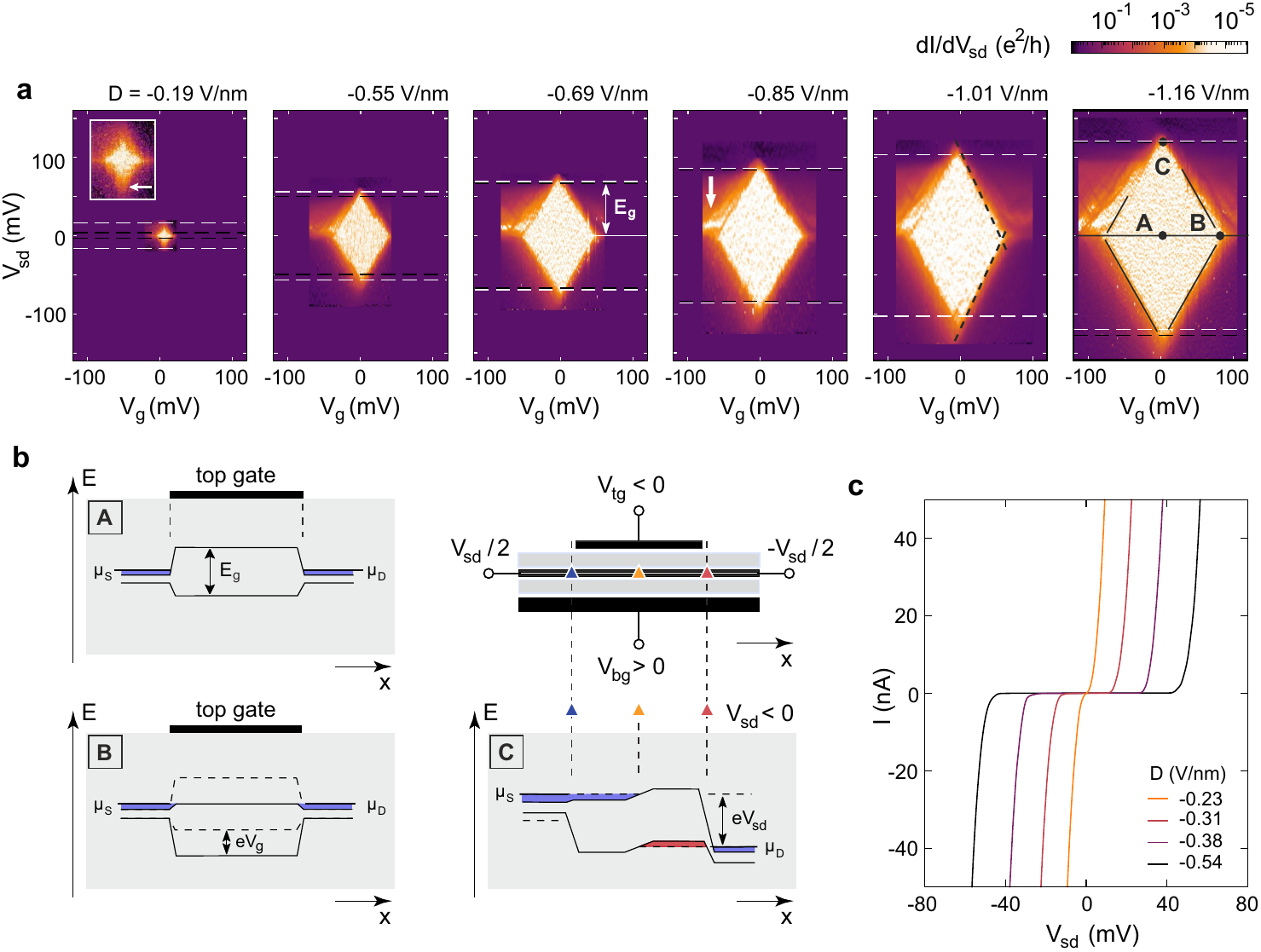}
	\caption{
		\textbf{(a)} Color plot of the differential conductance $\mathrm{d}I/\mathrm{d}V_\mathrm{sd}$ of a Gr/hBN/BLG device, measured at $T = 50~$mK
		as a function of $V_\mathrm{sd}$ and $V_\mathrm{g}$ for different  displacement fields (see labels). The white dashed-lines denote the band gap predicted by the model of ~\cite{McCann_2013} (for $\varepsilon_\text{z}=\overline{\varepsilon}=\varepsilon_\mathrm{BLG}=2$), the black dashed lines denote the effective gap $E_\mathrm{g}^\mathrm{eff}$ extracted as described in the main text. The inset shows a magnification for $D=-0.19$~V/nm. Even at small displacement fields, the area of suppressed differential conductance presents a pronounced diamond shape. \textbf{(b)}	Schematic representation of the various transport regimes at the points labeled as A, B, C in the rightmost diamond in panel (a). At $V_\mathrm{sd}=0$ and V$_\text{g}=0$, transport is strongly suppressed by the presence of a band gap in the double-gated region (A). Transport is re-established either by changing the chemical potential in the double-gated region using the effective gate voltage $V_\mathrm{g}$ (B) or by applying a sufficiently large source-drain voltage $V_\mathrm{sd}$, which changes the effective potential at the edges of the double-gated BLG, introducing charge carriers and forming a p-n junction.
    \textbf{(c)} I-V-characteristic of the Gr/hBN/BLG device for different values of $D$ at constant $V_\mathrm{g}$. The device shows a clear diode-like behavior, with no appreciable sub-threshold current.
	}
	\label{fig:exp_04}
\end{figure*}

\section{Direct observation of the band gap}
\label{Finite-bias spectroscopy}

Measurements like those presented in Figures~3b and~3c allow to extract the size of the electrostatically induced band gap, $E_\mathrm{g}$, but not to observe it directly.
To directly probe the band gap, we use finite-bias spectroscopy at low temperatures, i.e., we
measure the differential conductance of our devices, $g=\mathrm{d}I/\mathrm{d}V_\mathrm{sd}$, as a function of $V_\mathrm{g}$ and of $V_\mathrm{sd}$. This is a sensitive method to probe the characteristic energy scales of a system and the presence of localized states.

Figure~\ref{fig:exp_04}a shows finite-bias spectroscopy measurements performed at $T=50$~mK in a device with a graphite gate, for different values of the displacement field $D$. Diamond-shaped regions of strongly suppressed conductance can be observed around $V_\mathrm{g}=V_\mathrm{sd}=0$, where the size of the diamond scales with the applied $D$-field. For large values of $D$, the extent of the region of strongly suppressed conductance along the $V_\mathrm{sd}$ axis agrees very well with the size of the band gap predicted by theory, represented here by the horizontal white dashed lines.

The schematics of Figure~\ref{fig:exp_04}b depict different characteristic regimes that correspond to the points marked with A, B, C in the rightmost panel of Figure~\ref{fig:exp_04}a. Point A represents the condition of $V_\mathrm{sd}=V_\mathrm{g}=0$ at the center of the diamond, where charge transport through the device is suppressed the most. Point B represents the onset of charge transport induced by tuning $V_\mathrm{g}$ such that the conduction band in the double-gated area of the device is aligned with the conduction band of the leads. Point C represents the onset of conduction caused by a sufficiently large source-drain bias voltage applied symmetrically over the gated region (i.e.,
$\pm V_\mathrm{sd}/2=\pm E_\mathrm{g}/(2e)$).
The source-drain voltage creates a p-n junction within the BLG, over which the complete $V_\mathrm{sd}$ drops as long as there is no current flowing (for more details see Supplementary Material). This results in the diode-like behavior of the bias-dependent current shown in Figure~\ref{fig:exp_04}c, where the threshold voltage clearly depends on the $D$-field and therefore on the size of the band gap.\\

Please note that the slight asymmetries in $\mathrm{d}I/\mathrm{d}V_\mathrm{sd}$ outside the diamonds with respect to V$_\text{sd}$ and $V_\text{g}$ (see e.g. white arrow in the 4th panel of Figure~4a) can be explained by the $D$-field dependent layer polarization and a spatially varying disorder potential.
For example, for negative displacement field the carriers near the valance band edge ($V_\text{g}<0$) are located in the upper BLG layer~\cite{McCann_2013,Levitov2017} (see Fig. 1b), which is closer to the Au top gate. Thus, the asymmetry in $V_\text{g}$ indicates asymmetries in the residual disorder at the Au top and graphite bottom gate interface.
The asymmetry in V$_\text{sd}$ indicates a spatial variation of disorder, as the location of the p-n junction underneath the top gate in this case depends on the polarity of V$_\text{sd}$ (for more details, see Supplementary Material).
\begin{figure*}[t]
	\includegraphics[width=0.95\linewidth]{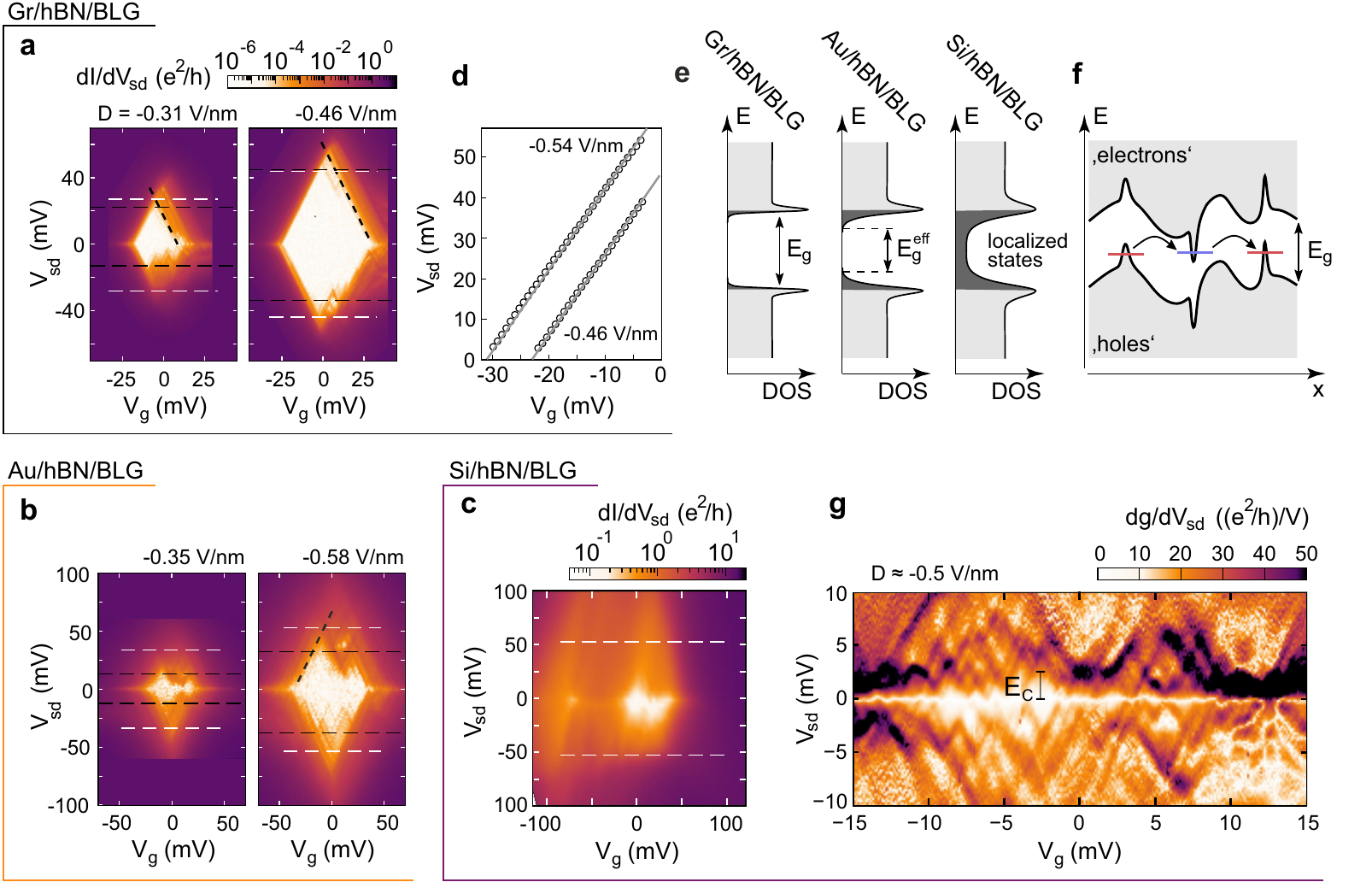}
	\caption{
\textbf{(a, b)} Color plots of the differential conductance of a (second) Gr/hBN/BLG device (a) and of a Au/hBN/BLG device (b), measured as a function of  $V_\mathrm{sd}$ and $V_\mathrm{g}$ at 1.6~K for different displacement fields. As in Figure 4a, the dashed white lines indicate the size of the band gap as predicted by the model of Ref.~\cite{McCann_2013}; the horizontal black-dashed lines the size of the effective gap, $E_\mathrm{g}^\mathrm{eff}$, extracted as discussed in the main text. The diagonal black-dashed lines have a slope of 2.
\textbf{(c)} Differential conductance of a Si/hBN/BLG device, measured at 1.6~K and $D=-0.5$~V/nm. No clear feature related to the band gap can be observed.
\textbf{(d)} Extracted diamond edge (threshold resistance $10^9$~$\Omega$) for two different $D$-fields (see labels). The gray lines have a slope of exactly 2.
\textbf{(e)} Schematics of the density of states (DOS) in the three different types of BLG devices. Disorder-induced localized states give rise to tail states with subgap energies that reduce or effectively close the band gap.
\textbf{(f)} Schematic representation of the band gap of a Si/hBN/BLG device. The presence of strong potential disorder gives rise to localized states with subgap energies, leading to hopping transport through the band gap.
\textbf{(g)} Derivative of the differential conductance $g=\mathrm{d}I/\mathrm{d}V_\mathrm{sd}$ for small $V_\mathrm{sd}$. The diamond-like features can be associated to localized electronic states or charge islands with charging energy $E_\mathrm{C}\approx 2-3$~meV.
	}
	\label{fig:exp_05}
\end{figure*}

Figures~\ref{fig:exp_05}a-c present the same type of bias-spectroscopy measurements performed at $T=1.6$~K on a second device with a graphite bottom gate (Figure~\ref{fig:exp_05}a), as well as on a device with gold bottom gate (Figure~\ref{fig:exp_05}b) and on one with silicon bottom gate (Figure~\ref{fig:exp_05}c). Also at this higher temperature, the device with graphite  bottom gate shows at large $D$-fields a well-defined diamond-shaped region of strongly suppressed conductance around $V_\mathrm{g}=V_\mathrm{sd}=0$, whose span along the $V_\mathrm{sd}$ axis is in fairly good agreement with the size of the band gap predicted by theory (horizontal white dashed lines).

The slopes of the diamond outlines in the $V_\mathrm{sd}-V_\mathrm{g}$ map is very close to 2 as highlighted by the dashed lines (see also the second rightmost panel in Figure~\ref{fig:exp_04}a).
In Figure~\ref{fig:exp_05}d we show the extracted outline of the diamond (resistance threshold value of $10^9$~$\Omega$) for two different displacement fields (see labels). In both cases, the slope agrees very well with a slope of 2 (see gray lines).
This means that $V_\mathrm{g}$ indeed tunes directly the chemical potential $\mu=eV_\mathrm{g}$ in the band gap region and no trap states need to be charged.
The slope of 2 results from the fact that when starting in configuration A (see Figures~4a,b) we need either $eV_\mathrm{sd}=E_\mathrm{g}$ to lift the band gap (moving to point C) or $\mu=E_\mathrm{g}/2 = eV_\mathrm{g}$ (moving to point B) resulting in $\Delta V_\mathrm{sd} = 2 \Delta V_\mathrm{g}$.

The fact that this can be observed, in turn, unambiguously indicates the absence of impurity bands, localized states or any other trap state that can potentially be charged in the device.

The device with a gold bottom gate presents also a region of strongly suppressed conductance for large $D$-fields. However, this does not appear as a single, well-defined diamond,
but as a series of overlapping diamonds with different sizes and positions  (see right panel of Figure~\ref{fig:exp_05}b). The device with silicon bottom gate shows overall much higher conductance, with no clear feature that can be related to the band gap energy (Figure~5c).

As already discussed in Section II, the different behavior of the three devices can be traced back to their different level of disorder.
In the device with a silicon bottom gate, the unscreened potential of charges trapped in the SiO$_2$ substrate or at the SiO$_2$/hBN interfaces creates localized states with subgap energies~\cite{Ando2006,Ishigami2007,Chen2008Mar,Morozov2008,Zhang2009Aug,Dean2010}, as sketched in Figures~\ref{fig:exp_05}e,f. Signatures of hopping transport  
can indeed be found in the derivative of the differential conductance $d\mathrm{g}/dV_\mathrm{sd}$, see Figure~\ref{fig:exp_05}g.
Here, we observe diamond-like features associated to the charging of charge islands or of individual localized states, as characteristic for statistical Coulomb blockade~\cite{Stampfer2009Feb,Gallagher2010Mar}. The extent of these diamond features in $V_\mathrm{sd}$ is related to the characteristic charging energy $E_\mathrm{C}$ of a localized state, a charge puddle, or to the energy spacing between different states. Here, we find typical energies in the range of 2-3~meV. The observation of individual charging events indicates that the total number of localized states (or puddles) contributing to transport is rather limited. Nevertheless, these states provide a percolation channel through the gated BLG region and prevent a complete current suppression, in good agreement with earlier reports~\cite{Zou2010Aug,Taychatanapat_PhysRevLett2010,Kanayama2015,Sui_nature2015}.

The presence of a local bottom gate -- either of gold or graphite --  strongly screens the disorder potential caused by charged impurities and allows to open a real band gap in BLG by means of a displacement field. However, the band gap is ultraclean only in the case of devices with a graphite gate, while the device with a gold gate still presents signatures of transport through localized states with subgap energies.
The presence of these localized states results in tail states in the density of states (DOS), which reduce or -- in the case of silicon-gated devices -- effectively suppress the band gap (see schematics in Figure~\ref{fig:exp_05}e).

To quantitatively compare the band gaps observed by transport spectroscopy with theory, we define the effective band gap $E^{\mathrm{eff}}_\mathrm{g}=e\, (V^+_\mathrm{sd,th}+|V^-_\mathrm{sd,th}|)/2$, where $V^{\pm}_\mathrm{sd,th}$ are the values of $V_\mathrm{sd}$ at which the differential conductance is equal to $(\mathrm{d}I/\mathrm{d}V_\mathrm{sd})_\mathrm{th}=10^{-9}$~S at $V_\mathrm{g}=0$.
These threshold voltages are indicated as dashed black lines in Figure~\ref{fig:exp_04}a and Figures~\ref{fig:exp_05}a,b.
Note that the differential conductance of the Si/hBN/BLG device is always higher than $10^{-9}$~S. Figure~6 shows the values of the effective gap $E^{\mathrm{eff}}_\mathrm{g}$ as a function of the displacement field $D$ for two devices with a graphite bottom gate and for one with a gold bottom gate. For the devices with a graphite bottom gate, we observe an excellent agreement between the band gap measured by transport spectroscopy, $E^{\mathrm{eff}}_\mathrm{g}$, and the values predicted by theory for $D$-fields larger than $0.3$~V/nm.
The deviations at smaller values of $D$ indicate the presence of some residual disorder, whose influence is larger for small band gaps (see also inset of the leftmost panel in Figure~4a).
Vice versa, for the device with a gold bottom gate the effective transport gap $E^{\mathrm{eff}}_\mathrm{g}$ is around $20$~meV smaller compared to the expected values.
This confirms the presence of extended tail states within the band gap, as sketched in Figure~\ref{fig:exp_05}e.
As for all devices the preparation of the BLG has been the same, rather than impurities or defects related to the BLG, these tail states are most likely induced
by substrate roughness, or by contaminations at the interfaces due to the fabrication process of the gold-gated device~\cite{Zibrov2017,Rhodes2019,Yankowitz2019} or
by the disorder caused by grain boundaries in the gold~\cite{Osvald1992} and are not caused by impurities or edge states in the BLG itself.

\begin{figure}[t!]
	\centering
	\includegraphics[width=1.0\linewidth]{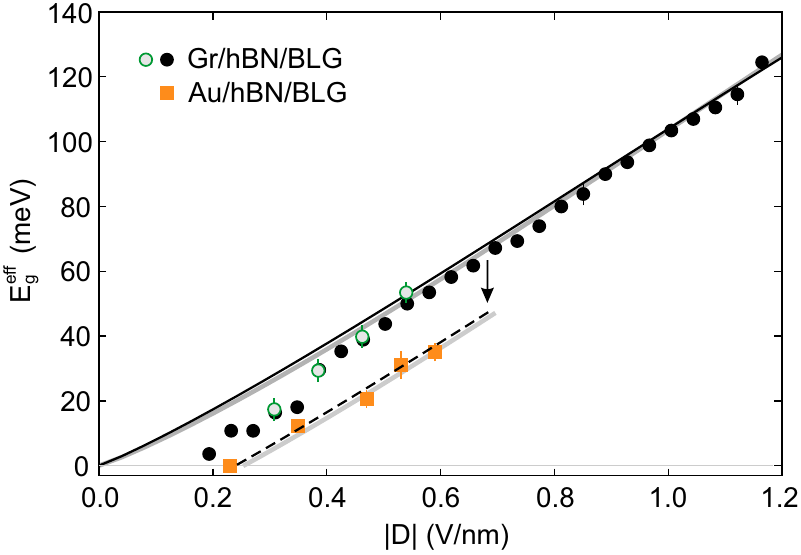}
	\caption{Comparison between the effective values of the band gap $E_\mathrm{g}^\mathrm{eff}$ extracted by transport spectroscopy measurements and the values of $E_\mathrm{g}$ predicted by theory (black line: model of Ref.~\cite{McCann_2013}, gray line: simplified model of~Ref.~\cite{Slizovskiy2021} with $\varepsilon_\text{z}$=1.65). The overall good agreement between theory and the value of  $E_\mathrm{g}^\mathrm{eff}$ extracted from the Gr/hBN/BLG devices (black circles from device \#1 (Figure~\ref{fig:exp_04}a); green circles from device \#2 (Figures~\ref{fig:exp_05}a,d)) indicate that these devices are only affected by very weak residual potential-disorder. Vice versa, the discrepancy observed for the Au/hBN/BLG device (orange squares) indicates the presence of disorder-induced tail states, as sketched in Figure~5e.}
	\label{fig:exp_3}
\end{figure}

\section{Summary and outlook}

To conclude, we showed that finite-bias spectroscopy is a versatile method to characterize the band gap in BLG. Its high sensitivity allows comparing the influence of (electrostatic) disorder potentials for different gating-technologies.
The measurements clearly indicate that devices with a graphite bottom gate fabricated as part of the van-der-Waals heterostructure outperform devices with gold and Si/SiO$_2$ gates, and behave very closely to what theory predicts for ideal BLG. In graphite-gated devices, we achieve band gaps of about $120~$meV with resistances up to 100~G$\Omega$ within the band gap (only limited by the setup). These results underline the importance of graphite as a bottom gate for BLG-based van-der-Waals heterostructures.

The high quality of Gr/hBN/BLG devices demonstrated in this work allows to readdress the broad field of possible applications offered by BLG, such as diodes~\cite{Shioya2012Jan}, phonon-lasers~\cite{Kuzmenko2009Nov,Tang2010Nov}, hot-electron bolometers~\cite{Yan2012Jul}, field-effect transistors (FET)~\cite{Cheli2009,Das2011Jan,Sako2011Aug} and tunnel FETs~\cite{Alymov2016}. The latter two are especially interesting for terahertz (THz) detection, where graphene and BLG based devices have already shown promising results~\cite{Dhillon2017, Bandurin2018}. The low disorder of Gr/hBN/BLG technology offers the possibility of significantly improving the device performance for this type of application, e.g., as recently demonstrated for tunnel FETs used for THz detection~\cite{Gayduchenko2021}.

Other applications that can greatly profit from the excellent tunability of the band gap in Gr/hBN/BLG devices are those based on proximity-induced properties in BLG, such as superconductivity~\cite{Kraft2018April,Li2020May}, exchange coupling~\cite{Zollner2020Nov} or strong spin-orbit coupling (SOC)~\cite{Levitov2017}. For example, by placing a strong SOC material on top of BLG, only the band associated with the layer close to the SOC material will exhibit spin splitting, thanks to the layer dependence of the bands close to the K-point (see colors in Figure~1).
This makes it possible to switch on and off the spin-orbit interaction by simply switching the sign of the $D$-field~\cite{Levitov2017}. Such an effect has been indeed recently demonstrated by taking advantage of the low disorder and the excellent control of the band gap with the applied $D$-field in BLG devices with graphite gates~\cite{Island2019}, which are therefore an  interesting platform for spin-orbit valves and spin transistors~\cite{Gmitra2017Oct}.

Furthermore, Gr/hBN/BLG technology has allowed realizing sophisticated devices such as quantum point contacts~\cite{Overweg2018Jan,Kraft2018Dec,Banszerus2020May,Lee2020Mar}, and quantum dots in BLG with single-electron control~\cite{Eich2018Jul,Banszerus2018Aug,Banszerus2020Feb}.
This underlines the possibility of BLG as a potential host material for spin and valley qubits~\cite{Wu2013}. Moreover, thanks to a long electron phase coherence length~\cite{Engels2014Sep} it promises to be an interesting platform for mesoscopic physics in low-dimensions.

In short, our study unambiguously shows that the Gr/hBN/BLG technology allows realizing van-der-Waals heterostructures that truly behave as  semiconductors with an electrostatically tunable gap.
This opens up a wide field of possible applications, especially when considering that a required scalability can also be enabled by using high quality CVD BLG material~\cite{Liu2012Sep,Schmitz2017Jun,Bisswanger2022Jun}.
\newline
\newline

\textbf{Acknowledgements} The authors thank S.~Trellenkamp, F.~Lentz, and D.~Neumaier for their support in device fabrication and F.~Hassler, D.~Kennes, A.~Garcia-Ruiz  and F.~Haupt for discussions.
This project has received funding from the European Union's Horizon 2020 research and innovation program under grant agreement No. 881603 (Graphene Flagship) and from the European Research Council (ERC) under grant agreement No. 820254, the Deutsche Forschungsgemeinschaft (DFG, German Research Foundation) under Germany's Excellence Strategy - Cluster of Excellence Matter and Light for Quantum Computing (ML4Q) EXC 2004/1 - 390534769, through DFG (BE 2441/9-1 and STA 1146/11-1), and by the Helmholtz Nano Facility~\cite{Albrecht2017May}.  K.W. and T.T. acknowledge support from the Elemental Strategy Initiative conducted by the MEXT, Japan (Grant Number JPMXP0112101001) and JSPS KAKENHI (Grant Numbers 19H05790, 20H00354 and 21H05233).

\textbf{Data availability}
The data supporting the findings are available in a
Zenodo repository under accession code 10.5281/zenodo.6119509.

\bibliographystyle{adms}

\bibliography{literature.bib}

\end{document}


\title{\bf Supplemental Material for: \\ Direct observation of ultraclean tunable band gaps in bilayer graphene}

\author{E.~Icking}
\affiliation{JARA-FIT and 2nd Institute of Physics, RWTH Aachen University, 52074 Aachen, Germany,~EU}
\affiliation{Peter Gr\"unberg Institute  (PGI-9), Forschungszentrum J\"ulich, 52425 J\"ulich,~Germany,~EU}
\author{L.~Banszerus}
\affiliation{JARA-FIT and 2nd Institute of Physics, RWTH Aachen University, 52074 Aachen, Germany,~EU}
\affiliation{Peter Gr\"unberg Institute  (PGI-9), Forschungszentrum J\"ulich, 52425 J\"ulich,~Germany,~EU}
\author{F. W\"ortche}
\affiliation{JARA-FIT and 2nd Institute of Physics, RWTH Aachen University, 52074 Aachen, Germany,~EU}
\author{F.~Volmer}
\affiliation{JARA-FIT and 2nd Institute of Physics, RWTH Aachen University, 52074 Aachen, Germany,~EU}
\author{P.~Schmidt}
\affiliation{JARA-FIT and 2nd Institute of Physics, RWTH Aachen University, 52074 Aachen, Germany,~EU}
\affiliation{Peter Gr\"unberg Institute  (PGI-9), Forschungszentrum J\"ulich, 52425 J\"ulich,~Germany,~EU}
\author{C.~Steiner}
\affiliation{JARA-FIT and 2nd Institute of Physics, RWTH Aachen University, 52074 Aachen, Germany,~EU}
\affiliation{Peter Gr\"unberg Institute  (PGI-9), Forschungszentrum J\"ulich, 52425 J\"ulich,~Germany,~EU}
\author{S.~Engels}
\affiliation{JARA-FIT and 2nd Institute of Physics, RWTH Aachen University, 52074 Aachen, Germany,~EU}
\affiliation{Peter Gr\"unberg Institute  (PGI-9), Forschungszentrum J\"ulich, 52425 J\"ulich,~Germany,~EU}
\author{J.~Hesselmann}
\affiliation{JARA-FIT and 2nd Institute of Physics, RWTH Aachen University, 52074 Aachen, Germany,~EU}
\author{M.~Goldsche}
\affiliation{JARA-FIT and 2nd Institute of Physics, RWTH Aachen University, 52074 Aachen, Germany,~EU}
\affiliation{Peter Gr\"unberg Institute  (PGI-9), Forschungszentrum J\"ulich, 52425 J\"ulich,~Germany,~EU}
\author{K.~Watanabe}
\affiliation{
National Institute for Materials Science, 1-1 Namiki, Tsukuba, 305-0044, Japan }
\author{T.~Taniguchi}
\affiliation{
International Center for Materials Nanoarchitectonics, National Institute for Materials Science, 1-1 Namiki, Tsukuba 305-0044, Japan}
\author{C.~Volk}
\affiliation{JARA-FIT and 2nd Institute of Physics, RWTH Aachen University, 52074 Aachen, Germany,~EU}%
\affiliation{Peter Gr\"unberg Institute  (PGI-9), Forschungszentrum J\"ulich, 52425 J\"ulich,~Germany,~EU}
\author{B.~Beschoten}
\affiliation{JARA-FIT and 2nd Institute of Physics, RWTH Aachen University, 52074 Aachen, Germany,~EU}%
\author{C.~Stampfer}
\affiliation{JARA-FIT and 2nd Institute of Physics, RWTH Aachen University, 52074 Aachen, Germany,~EU}%
\affiliation{Peter Gr\"unberg Institute  (PGI-9), Forschungszentrum J\"ulich, 52425 J\"ulich,~Germany,~EU}%

\maketitle

\section{Methods}
\subsection{Fabrication of gated BLG devices}\label{Fabrication}
State-of-the-art bilayer graphene (BLG) devices used in transport experiments are fabricated by the well established van-der-Waals dry pick-up technique~\cite{Wang2013Nov,Engels2014Sep}, where different exfoliated crystals are subsequently picked up with a thin membrane of polycarbonate ~\cite{Wang2013Nov, Purdie_2018}.
In our work, we consider the three classes of devices shown in Figure~2a of the main manuscript, which differ only on the type of bottom gate. In all devices, an exfoliated BLG flake is encapsulated between two 20 to 30~nm thick crystals of hexagonal boron nitride (hBN).

\subsubsection{Gr/hBN/BLG devices}
To fabricate the devices with graphite (Gr) bottom-gate, we use the dry assembly technique~\cite{Wang2013Nov,Engels2014Sep} to pick up one after the other the following crystals: exfoliated hBN ($20-30$~nm thick), exfoliated BLG, exfoliated hBN ($20-30$~nm thick), and exfoliated graphite ($5-15$~nm thick). The finished stack is transferred onto a Si$^{++}$/SiO$_2$ substrate for further processing (e-beam lithography and metal e-beam evaporation followed by lift-off for the metallic electrodes).

The structural ``quality" of the BLG is verified by scanning confocal Raman spectroscopy~\cite{Graf2007Feb,Neumann2015Sep}, which provides essential insights on its flatness and thus serves as a non-invasive quality check prior to further device processing~\cite{Ferrari2006,Yan2008,Cheng2015,Schmitz2017Jun}.

Ohmic contacts are fabricated by electron beam lithography (EBL), reactive ion etching (CF$_4$ plasma), and metal evaporation, closely following Ref.~\cite{Wang2013Nov}. In this type of device, the graphite crystal is contacted to serve as bottom gate.
In the very last step, we fabricate a Cr/Au top gate using EBL and e-beam evaporation of the metallic electrodes and lift-off.

\subsubsection{Au/hBN/BLG devices}
The fabrication of this type of devices is consistently more cumbersome than the previous one.
In this case, we use the dry-assembly technique~\cite{Wang2013Nov,Engels2014Sep} to form a hBN/BLG/hBN stack, which we deposit on a silicon substrate. We then use EBL, metal evaporation, and lift-off to form a Cr/Au gate. The entire stack -- including the Au gate -- is then flipped using a stack-flipping process similar to the one described in Ref.~\cite{zeng2019Apr}. After the flipping, the Au gate serves as a bottom gate. This stack-flipping approach ensures a homogeneous and flat interface between hBN and Au gate. As the next steps, we fabricate ohmic contacts and a Cr/Au top gate, as described for the Gr/hBN/BLG devices.

\subsubsection{Si/hBN/BLG devices}
To fabricate the devices with a highly doped silicon (Si) back-gate, we repeat all the steps described for the Gr/hBN/BLG devices, except for the picking-up of the graphite crystal, i.e., in this case we place an hBN/BLG/hBN stack on a highly doped Si$^{++}$-substrate covered with 285~nm SiO$_2$, and use the highly doped Si-substrate as a bottom gate.


\section{Resistance maps and quantum Hall measurements of all investigated devices}
\begin{figure}[!h]
\centering

\includegraphics[draft=false,keepaspectratio=true,clip,width=0.6\linewidth]{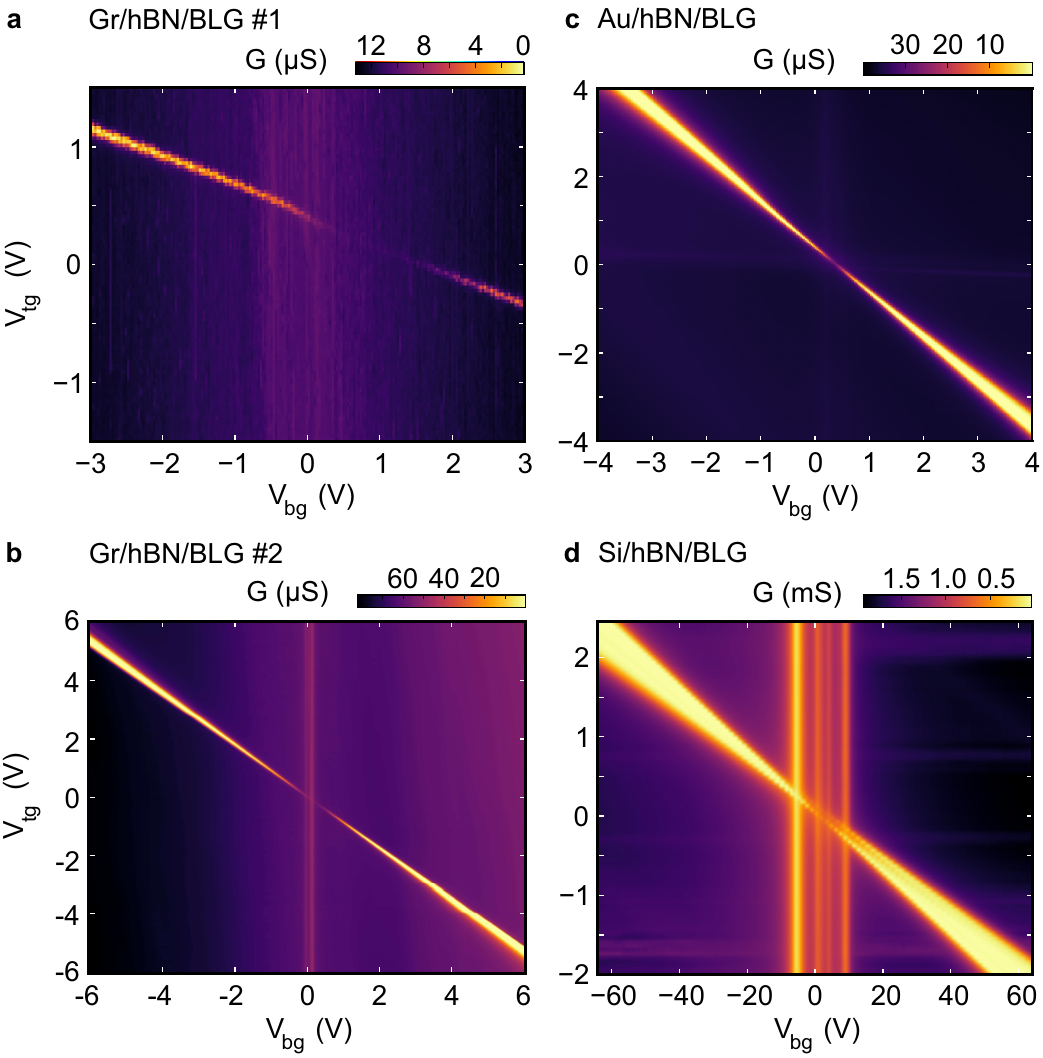}
\caption[FigS1]{Two-terminal conductance as a function of the top gate voltage $V_\mathrm{tg}$ and the bottom gate voltage $V_\mathrm{bg}$ to extract the relative lever arm for \textbf{(a)} the Gr/hBN/BLG \#1 device in Figure~4a at a constant bias voltage $V_\mathrm{sd}=0.1\;\text{mV}$ and $T=50$~mK, \textbf{(b)} the Gr/hBN/BLG \#2 device in Figure~5a at $V_\mathrm{sd}=1\;\text{mV}$ and $T=1.6$~K, \textbf{(c)} the Au/hBN/BLG device in Figure~5b at $V_\mathrm{sd}=1\;\text{mV}$ and \textbf{(d)} the Si/hBN/BLG device in Figure~5c at $V_\mathrm{sd}=0.1\;\text{mV}$.} \label{fig:s1x}
\end{figure}

\begin{figure}[!htp]
\centering

\includegraphics[draft=false,keepaspectratio=true,clip,width=0.6\linewidth]{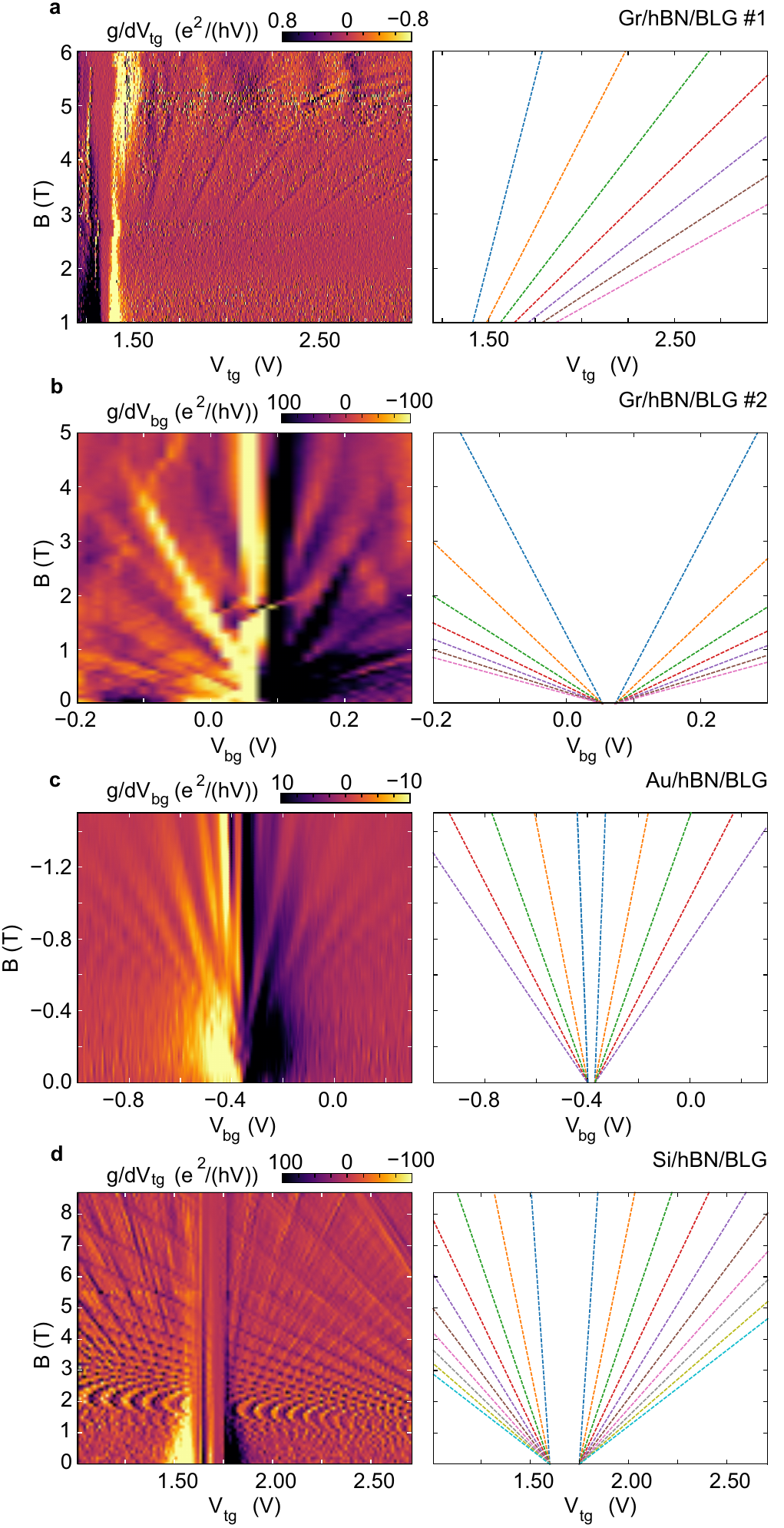}
\caption[FigS1]{Two-terminal quantum Hall measurements to extract the gate lever arm for \textbf{(a)} the Gr/hBN/BLG \#1 device used to measure the data shown in Figures~2f and~4a at a constant bias voltage $V_\mathrm{sd}=0.1\;\text{mV}$ and $T=50$~mK, \textbf{(b)} the Gr/hBN/BLG \#2 device used for the data presented in Figure~5a at $V_\mathrm{sd}=1\;\text{mV}$ and $T=1.6$~K, \textbf{(c)} the Au/hBN/BLG device (Figures~2d,e and 5b) at $V_\mathrm{sd}=1\;\text{mV}$ and \textbf{(f)} the Si/hBN/BLG device (Figures~2e and~5c) at $V_\mathrm{sd}=0.1\;\text{mV}$. Using the extracted lever-arm and equation~(S1) the Landau level spectra is depicted on the right side.
}
\label{fig:r3}
\end{figure}

Here we present the resistance-map measurements and the quantum Hall measurements performed on all investigated devices, see Figure~\ref{fig:s1x} and \ref{fig:r3}. All measurements are performed at $1.6$~K, except otherwise stated.

To record the resistance maps, we apply a small source-drain bias  $V_\mathrm{sd}=1$~mV and measure the current as a function of the voltage applied to the top and bottom gate, $V_\mathrm{tg}$ and $V_\mathrm{bg}$, respectively. From the offset of the charge neutrality point from $V_\mathrm{tg}=V_\mathrm{bg}=0$, we determine the values of $V_\mathrm{bg}^0$ and $V_\mathrm{tg}^0$ that enter in Equation~(3) and Equation~(4) in the main manuscript.

\begin{table}[h]
    \centering
    \renewcommand{\thetable}{S\arabic{table}}
    \begin{tabular}{|c|c|c|}
    \hline
      Device  & $V_\mathrm{tg}^0$ (mV) & $V_\mathrm{bg}^0$ (mV) \\ \hline
       Gr/hBN/BLG \#1  &  36  & 1020   \\ \hline
       Gr/hBN/BLG \#2 & 197  & -190  \\ \hline
       Au/hBN/BLG  & 41  & 331 \\ \hline
        Si/hBN/BLG & -80  & 2250  \\ \hline
    \end{tabular}
    \caption{Offset of the charge neutrality point with respect to $V_\mathrm{tg}=V_\mathrm{bg}=0$.}
    \label{tab1}
\end{table}
Here and in the following, Gr/hBN/BLG \#1 and Gr/hBN/BLG \#2 indicate the two different graphite-gated devices that have been measured in this work. All data labeled as Gr/hBN/BLG in the main manuscript have been measured on device  Gr/hBN/BLG \#1, except for the data shown in Figure~5a, which have been measured on Gr/hBN/BLG \#2.\\

We use quantum Hall measurements to extract the lever arm of the top- and bottom gate, $\alpha_\mathrm{tg}$ and $\alpha_\mathrm{bg}$ to the BLG. This is done by fitting the position of the Landau levels in plots like those shown in Figure~\ref{fig:r3} with the equation
\begin{equation}
    B=\frac{h}{\nu e}\alpha_\mathrm{tg(bg)}V_\mathrm{tg(bg)}+\mathrm{const.},
\end{equation}
where $\nu$ is the filling factor of the Landau level, $h$ the Planck constant, and $B$ the applied magnetic field~\cite{Zhao2010,Dauber2017,SonntagMay2018,Schmitz2020}. The values we extract for our devices are summarized in Table~S2.\\

Furthermore, we determine a residual charge carrier density of
$n_0= \alpha_\mathrm{bg} V_\mathrm{bg}^\mathrm{0} + \alpha_\mathrm{tg} V_\mathrm{tg}^\mathrm{0}$ for each device (see Table~S2).

\begin{table}[h]
    \centering
    \renewcommand{\thetable}{S\arabic{table}}
    \begin{tabular}{|c|c|c|c|c|}
    \hline
      Device & $\beta = \alpha_\mathrm{tg}/\alpha_\mathrm{bg}$ & $\alpha_\mathrm{bg}$ ($10^{11}$ cm$^{-2}$V$^{-1}$) & $\alpha_\mathrm{tg}$ ($10^{11}$cm$^{-2}$V$^{-1}$) & $n_0$ ( $10^{11}$cm$^{-2}$) \\ \hline
       Gr/hBN/BLG \#1  & 4.35 & 0.75$^*$  & 3.25  & 0.88 \\ \hline
        Gr/hBN/BLG \#2 & 0.89 & 5.72  & 4.4$^*$  &0.29 \\ \hline
Au/hBN/BLG & 0.99 &8.73 & 8.81$^*$  & $5.5$\\ \hline
        Si/hBN/BLG & 31.7 & 0.65$^*$  & 20.6  & $-0.19$\\ \hline
    \end{tabular}
    \caption{Gate lever-arms of top and bottom gate and the residual charge carrier density for each device shown in the main text. Values marked with a "$*$" are calculated using $\beta$.}
    \label{tab1}
\end{table}

\newpage

\section{Carrier Mobilities}
We extracted the carrier mobilities $\mu$ (at $T=1.6$K) from the resistance maps by fitting $R=R_c+(L/W)/(n e\mu)$, with the length $L$ and width $W$ of the double gated region, to the trace through the charge neutrality point along one of the gate axis, assuming a serial contact resistance $R_c$. From the fit we obtained the carrier mobilities in Table~R1. The rise in device quality for the different gating technologies is -- as expected -- also reflected in the increase in carrier mobility.\\

\begin{table}[!h]
    \centering
    \renewcommand{\thetable}{R\arabic{table}}
    \begin{tabular}{|c|c|c|c|}
    \hline
      Device & Gr/hBN/BLG \#2 & Au/hBN/BLG & Si/hBN/BLG \\ \hline
              Carrier mobility (cm$^2$/(Vs)) & 259.000 & 125.000  & 26.000 \\ \hline
        Contact Resistance (k$\Omega$) & 3.15 & 4.13  & 1.20 \\
        \hline
         $L/W$  & 3.24 & 1.48  & 2.00 \\
        \hline
    \end{tabular}
    \caption{Extraced carrier mobilities for the different device geometries. The rise in mobility also reflects the rise in device quality.}
    \label{tab1}
\end{table}

\section{Electrostatics of double-gated bilayer graphene}
\label{ChemicalPotential}

\begin{figure}[!h]
	\centering
	\includegraphics[width=0.35\linewidth]{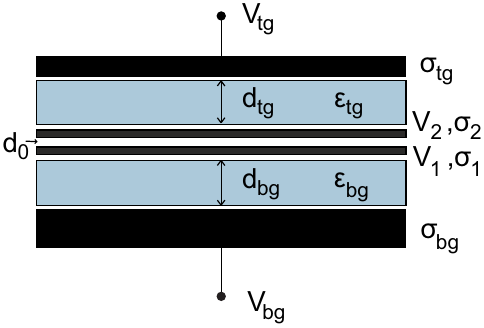}
	\caption{Schematic representation of dual gated bilayer graphene (BLG). Each gate is separated by a dielectric layer with dielectric constant $\varepsilon_\mathrm{tg}$, $\varepsilon_\mathrm{bg}$ and thickness $d_\mathrm{tg}$, $d_\mathrm{bg}$. The electric field produced by the top and bottom gate due to the applied voltages $V_\mathrm{bg}$ and $V_\mathrm{tg}$ induce potentials $V_1$ and $V_2$ on the bottom and top layer of BLG.
	}
	\label{fig:S0}
\end{figure}

The electrostatics of double-gated BLG can be described by two parallel conducting graphene plates separated by the BLG interlayer spacing of $d_0 \approx 0.34$~nm, which are placed between a bottom- and a top-gate (see Figure~\ref{fig:S0})~\cite{Cheli2009,Young2011,Alymov2016}. Each layer (1 and 2) of BLG is characterized by a potential $V_{1(2)}$ and a charge density $\sigma_{1(2)}=-e n_{1(2)}$, corresponding to the layer-dependent carrier densities $n_1$ and $n_2$. The total charge density on the BLG is $\sigma=\sigma_1+\sigma_2$.
The electric potentials $V_\mathrm{tg}$ and $V_\mathrm{bg}$ applied to the top and the bottom gate induce the potentials $V_{1(2)}$ on the layers of BLG~\cite{Alymov2016}. The difference between the potential on the two layers
\begin{equation}
\label{Delta1}
	\Delta=e(V_1-V_2),
\end{equation}
is responsible for the opening of a gap in the band structure of BLG (see Figure~1 in the main manuscript), whereas the averaged potential of the two layers gives the chemical potential of the BLG:
\begin{equation}
\label{mu}
	\mu=e\frac{V_1+V_2}{2}.
\end{equation}

To determine the relation between the layer potential $V_{1(2)}$ and $V_\mathrm{tg(bg)}$, we define the following capacities per unit area:  $C_\mathrm{tg}=\varepsilon_0\varepsilon_\mathrm{tg}/d_\mathrm{tg}$ between the top gate and the upper graphene layer,
$C_\mathrm{bg}=\varepsilon_0\varepsilon_\mathrm{bg}/d_\mathrm{bg}$ between the bottom gate and the lower graphene layer, and $C_\mathrm{BLG}=\varepsilon_0\varepsilon_\mathrm{BLG}/d_\mathrm{0}$ between the two graphene layers of BLG.
Applying Gauss' law to the surfaces
gives a set of equations that connect the potentials and the charge densities on each BLG layer and on the gate electrodes~\cite{Young2011}:
\begin{equation}
\label{eq:S1}
C_\mathrm{tg}(V_\mathrm{tg}-V_{2})=\sigma_\mathrm{tg},
\end{equation}
\begin{equation}
\label{eq:S2}
C_\mathrm{BLG}(V_2-V_1)=\sigma_\mathrm{tg}+\sigma_\mathrm{2},
\end{equation}
\begin{equation}
\label{eq:S3}
C_\mathrm{bg}(V_\mathrm{bg}-V_1)=\sigma_\mathrm{bg},
\end{equation}
\begin{equation}
\label{eq:S4}
C_\mathrm{BLG}(V_1-V_2)=\sigma_\mathrm{bg}+\sigma_\mathrm{1}.
\end{equation}
Using these equations and the fact that the system also has to fulfill charge conservation,
$\sigma_\mathrm{tg}+\sigma_1+\sigma_2+\sigma_\mathrm{bg}=0$,
results in the following expression for $\sigma_1$ and $\sigma_2$
\begin{eqnarray}
-\sigma_1 &= & \sigma_2 +\sigma_\mathrm{tg}+\sigma_\mathrm{bg} = -C_\mathrm{BLG}(V_1-V_2)+C_\mathrm{bg}(V_\mathrm{bg}-V_{1}), \label{eq:S8}
\\-\sigma_2 &= & \sigma_1 +\sigma_\mathrm{bg}+\sigma_\mathrm{tg} = \phantom{-} C_\mathrm{BLG}(V_1-V_2)+C_\mathrm{tg}(V_\mathrm{tg}-V_{2}), \label{eq:S9}
\end{eqnarray}
which in turn results in the following expressions for the layer potentials:
\begin{equation}
\label{V1}
	V_1=\frac{\sigma_1+\sigma_2+C_\mathrm{bg}V_\mathrm{bg}+C_\mathrm{tg}V_\mathrm{tg}+C_\mathrm{tg}C_\mathrm{BLG}^{-1}(\sigma_1+C_\mathrm{bg}V_\mathrm{bg})}{C_\mathrm{bg}C_\mathrm{tg}\left[C_\mathrm{BLG}^{-1}+C_\mathrm{bg}^{-1}+C_\mathrm{tg}^{-1}\right]},
\end{equation}
\begin{equation}
\label{V2}
	V_2=\frac{\sigma_1+\sigma_2+C_\mathrm{bg}V_\mathrm{bg}+C_\mathrm{tg}V_\mathrm{tg}+C_\mathrm{bg}C_\mathrm{BLG}^{-1}(\sigma_2+C_\mathrm{tg}V_\mathrm{tg})}{C_\mathrm{bg}C_\mathrm{tg}\left[C_\mathrm{BLG}^{-1}+C_\mathrm{bg}^{-1}+C_\mathrm{tg}^{-1}\right]}.
\end{equation}
The chemical potential of BLG can then be written as
\begin{equation}\label{eq:mu}
\mu =e\frac{2(\sigma_1+\sigma_2)+2\left(C_\mathrm{bg}V_\mathrm{bg}+C_\mathrm{tg}V_\mathrm{tg}\right)+C_\mathrm{BLG}^{-1}\left[C_\mathrm{tg}(\sigma_1+C_\mathrm{bg}V_\mathrm{bg})+C_\mathrm{bg}(\sigma_2+C_\mathrm{tg}V_\mathrm{tg})\right]}{2C_\mathrm{bg}C_\mathrm{tg}\left[C_\mathrm{BLG}^{-1}+C_\mathrm{bg}^{-1}+C_\mathrm{tg}^{-1}\right]}.
\end{equation}

This equation needs in general to be solved self-consistently, since the charge density on the layers, $\sigma_{1(2)}$, depends on
the asymmetry $\Delta$ (see discussion in Sec.~IV).
~\newline
~\newline
However, if the chemical potential is in the band gap ($-E_g/2<\mu<E_g/2$), the total charge density of BLG has to be zero ($\sigma_1+\sigma_2=0$). Taking into account that in a typical device $C_\mathrm{BLG}\gg C_\mathrm{tg},C_\mathrm{bg}$, Equation~(\ref{eq:mu}) becomes an analytical expression of the voltages $V_\mathrm{bg}$ and $V_\mathrm{tg}$:
\begin{equation}\label{eq:mu_gap}
\mu \approx
 e\frac{C_\mathrm{bg}V_\mathrm{bg}+C_\mathrm{tg}V_\mathrm{tg}}{C_\mathrm{bg}+C_\mathrm{tg}}=e\frac{V_\mathrm{bg}+\beta V_\mathrm{tg}}{1+\beta} = e V_\mathrm{g},
\end{equation}
where $\beta=C_\mathrm{tg}/C_\mathrm{bg}$ is the relative lever-arm between top and bottom gate.
~\newline
~\newline
To express the potential asymmetry $\Delta$ between the layers, we start by expressing the difference in charge carrier density between the upper and lower layer $\delta n=n_2-n_1=(\sigma_1-\sigma_2)/e$ using Equation~(\ref{eq:S8}) and Equation~(\ref{eq:S9}). This results in
\begin{eqnarray}
e \delta{n} &=& 2 C_\mathrm{BLG} (V_\mathrm{1}-V_\mathrm{2}) - C_\mathrm{bg} (V_\mathrm{bg}-V_\mathrm{1}) + C_\mathrm{tg} (V_\mathrm{tg}-V_\mathrm{2}) \nonumber \\
&=& 2 C_\mathrm{BLG} \left[ V_\mathrm{1} \left( 1 + \frac{C_\mathrm{bg}}{2 C_\mathrm{BLG}}\right) - V_\mathrm{2} \left( 1 + \frac{C_\mathrm{tg}}{2 C_\mathrm{BLG}}\right) \right] - 2 D,
\end{eqnarray}
where in the last step we introduced the average displacement field
\begin{equation}\label{eq:D-field}
D = \frac{C_\mathrm{bg} V_\mathrm{bg} - C_\mathrm{tg} V_\mathrm{tg}}{2}.
\end{equation}
Taking again into account that in a typical device $C_\mathrm{BLG}\gg C_\mathrm{tg},C_\mathrm{bg}$ we obtain
\begin{eqnarray}
e \delta{n} &\approx& 2 C_\mathrm{BLG} \left(V_\mathrm{1} - V_\mathrm{2} \right) - 2 D \\
&=& 2 C_\mathrm{BLG} \frac{\Delta}{e} - 2 D,
\end{eqnarray}
where in the last step we made use of Equation~(\ref{Delta1}). From this expression we finally obtain:
\begin{equation}\label{Delta2a}
\Delta = \frac{e D}{C_\mathrm{BLG}}+\frac{e^2 \delta n}{2 C_\mathrm{BLG}},
\end{equation}
which corresponds to Equation~(64) in Ref.~\cite{McCann_2013}.

\subsection{Independent control of effective gate potential and displacement field}

Changing the chemical potential $\mu$ while maintaining a constant band gap is important in order to probe the band gap. Although Equation~(\ref{eq:mu_gap}) provides an expression for the chemical potential as a function of gate voltages, it does not yet include the constraint of a constant displacement field (i.e. constant band gap).\\

To keep the overall band gap constant while going e.g. from point A to point B in the rightmost panel of Figure~4a, the displacement field $D$ must be kept constant.
Combining Equation~(3) of the main text with Equation~(\ref{eq:mu_gap}) yields an expression for the top gate voltage
\begin{equation}
    V_\mathrm{tg}-V^0_\mathrm{tg}=\frac{1+\beta}{2\beta}V_\mathrm{g}-\frac{D}{e\beta\alpha_\mathrm{bg}}
\end{equation}
and the back gate voltage
\begin{equation}
    V_\mathrm{bg}-V^0_\mathrm{bg}=\frac{1+\beta}{2}V_\mathrm{g}+\frac{D}{e\alpha_\mathrm{bg}}
\end{equation}
as a function of the effective gate potential $V_\mathrm{g}$ and displacement field $D$. These equations dictate how the gate voltages must be changed (starting from point A, where $\mu \equiv 0$) to shift the chemical potential by a desired amount $\mu = e V_\mathrm{g}$ for a constant $D$.


\section{Self-consistent calculations of the band gap}

Equation~(\ref{Delta2a}) expresses the onsite potential difference $\Delta$ in terms of the difference of charge-carrier density in the two layers of BLG, $\delta n=n_2-n_1$, which is itself a function of $\Delta$.
In fact, the carrier densities on the layers of BLG, $n_{1(2)}$, can be calculated using the sub-lattice ($\lambda=A,B$) amplitudes $\phi^b_{\lambda,2(1),k}$ of the four ($b=1 - 4$) spin and valley degenerate bands~\cite{McCann_2013,Jung2014Jan}.

These are determined by the eigenstates $(\phi^b_{A1,k},\phi^b_{B1,k},\phi^b_{A2,k},\phi^b_{B2,k})^T$ of the $4 \times 4$ Hamiltonian of BLG
\begin{equation}
\label{hamiltonian}
    H=\left(
    \begin{array}{c c c c}
         \Delta/2& v_0\pi^\dagger & -v_4\pi^\dagger  &-v_3\pi\\
         v_0\pi& \Delta/2+\Delta' & \gamma_1  &-v_4\pi^\dagger\\
         -v_4\pi& \gamma_1 & -\Delta/2+\Delta'  & v_0\pi^\dagger\\
         -v_3\pi^\dagger& -v_4\pi & v_0\pi  & -\Delta/2\\
    \end{array}
    \right),
\end{equation}
where $\pi\equiv \hbar(\xi k_x+ik_y)$, with $\mathbf{k}=(k_x,k_y)$ the electron wave vector in the valleys
 $\mathbf{K}_\xi=\xi\left(4\pi/(3a),0\right)$, and $\xi=\pm1$.
Furthermore, $v_i\equiv \sqrt{3} a \gamma_i//(2\hbar)$, where $\gamma_i$ are coupling parameters and $a$ the lattice spacing of a graphene sheet ($i=0,1,3,4$). Following Ref.~\cite{McCann_2013}, $\gamma_1= 0.381$~eV
is the interlayer coupling strength (see Figure~1a in the main manuscript), $\gamma_0=3.16$~eV is the intralayer nearest neighbour coupling strength, $\gamma_3=0.38$~eV is the skew interlayer coupling between A1 and B2 sites giving rise to trigonal warping and $\gamma_4=0.14$~eV is the interlayer coupling between dimer and non-dimer orbitals~\cite{McCann_2013,Jung2014Jan}. The constant $\Delta'=0.015$~meV accounts for the energy difference between dimer and non-dimer site, and $\Delta$ describes the asymmetry of the on-site potential energies on the two layers.\\

Assuming an undoped gapped BLG (i.e., Fermi level in the band gap), the sum over the band index $b$ in Equation~(\ref{n2}) reduces to the first two bands only ($b=1,2$) with the band gap between bands 1,2 and 3,4~\cite{Slizovskiy2021}.
Rewriting the Hamiltonian in Equation~(\ref{hamiltonian}) in terms of low-energy and the non-dimer components only results in an effective $2 \times 2$ Hamiltonian~\cite{McCann_2013}
\begin{equation}
H=\left(
\begin{array}{c c}
\Delta/2 & -\frac{\hbar^2}{2m^*} (k_x-ik_y)^2\\
\frac{\hbar^2}{2m^*} (k_x+ik_y)^2 & -\Delta/2
\end{array}\right),
\label{2xHamiltonian}
\end{equation}
with the effective mass $m^*= \gamma_1/2 v_\text{F}^2\approx0.033 \;m_\textbf{e}$, where $v_\text{F}$ is the Fermi velocity of graphene.

\label{SelfConsistentCalculations}
~\newline
~\newline
The onsite potential difference $\Delta$ in Equation~(2) in the main text depends on the difference in the charge carrier densities $\delta n = n_2 - n_1$ on the individual BLG layers~\cite{McCann_2013,Slizovskiy2021}. The difference can be nonzero even for a total charge carrier density of $n=n_2+n_1=0$. Since the individual carrier densities $n_{2(1)}$ themself depend on the potential difference $\Delta$~\cite{McCann_2013,Slizovskiy2021}, there is no analytical solution to this problem. However, there are two references, i.e., models, that present a self-consistent approach.
\subsection{Model by McCann and Koshino -- 2013}
Following McCann and Koshino~\cite{McCann2006Mar,McCann_2013}, by taking into account screening using the tight-binding model and the Hartree theory, the plate capacitor model leads (for $n=0$) to an on-site potential difference of (see also Equation~(\ref{Delta2a})):
\begin{equation}
\label{Delta}
    \Delta=\frac{ d_0 e D}{\varepsilon_0\varepsilon_\mathrm{BLG}}+\frac{d_0 e^2}{2\varepsilon_0\varepsilon_\mathrm{BLG}}\delta n
\end{equation}
with the dielectric constant $\varepsilon_\mathrm{BLG}$ of BLG and the difference in charge carrier densities between the upper and lower layer $\delta n=n_2-n_1$ expressed by~\cite{McCann_2013}:
\begin{equation}
\label{delta_n}
    \delta n = \frac{n_\perp\Delta}{2\gamma_1}\ln\left(\frac{\Delta}{4\gamma_1}\right).
\end{equation}
Here, the quantity $n_\perp = \gamma^2_1 / (\pi\hbar^2v_\mathrm{F}^2)$ describes a characteristic carrier density scale.
Note that within this screening model, the difference $\delta n$ in carrier density between the individual layers depends non-linearly on $\Delta$ and changes with the size of the band gap.
Substituting Equation~(\ref{delta_n}) into Equation~(\ref{Delta}) results in an expression for the on-site potential:
\begin{equation}
	\Delta\approx\Delta_\mathrm{0}(D)\left[1-\frac{d_0e^2\gamma_1}{4\pi\hbar^2v_\text{F}^2\varepsilon_0\varepsilon_\mathrm{BLG}}\ln\left(\frac{\Delta}{4\gamma_1}\right)\right]^{-1},
	\label{SelfConv}
\end{equation}
with $\Delta_0(D)= d_0 e D /(\varepsilon_0\varepsilon_\mathrm{BLG})$ being the onsite potential difference without screening~\cite{McCann_206,McCann_2013}. To calculate the band gap as a function of applied displacement field we compute Equation~(\ref{SelfConv}) with $\Delta=\Delta_0$ as a starting parameter. We then obtain a $\Delta$ with included screening effects, which we use again to calculate Equation~(\ref{SelfConv}). This process is repeated until the computation converges. The result of this self-consistent calculation with $\varepsilon_\mathrm{BLG}=2$ is the black line in Figures~3c and 6 in the main text.

\subsection{Model by Slizovskiy et al. -- 2021}
Slizovskiy et al. follows a similiar approach~\cite{Slizovskiy2021} describing the on-site potential difference:
\begin{equation}
\label{Delta2}
    \Delta=\frac{ d_0 e D}{\varepsilon_0\varepsilon_\text{z}}+\frac{d_0 e^2}{2\varepsilon_0}\frac{1+1/\varepsilon_\text{z}}{2}\delta n.
\end{equation}
Here, $\varepsilon_\text{z}$ is the effective out-of-plane dielectric susceptibility of BLG~\cite{Slizovskiy2021}.
The layer densities $n_{2(1)}$ depend on the amplitude $\phi^b_{\lambda,2(1),k}$ of the wavefunction, which can be found by solving the Hamiltonian, and can be described as~\cite{Slizovskiy2021}
\begin{equation}
\label{n2}
    n_{2(1)}=\int\frac{d^2 \mathbf{k}}{\pi^2}\sum_{b=1,2}\left[\sum_{\lambda=A,B}\left|\phi^b_{\lambda,2(1),k}\right|^2-\frac{1}{4}\right]
\end{equation}
for $n=0$. Here, $b$ denotes the band number, with the band gap between bands 1,2 and 3,4, $\lambda$ denotes the sub-lattice~\cite{Slizovskiy2021}.
For low-energy band gaps and considering only the non-dimer components, i.e., approximating $n_{2(1)}$ by solving Equation~(\ref{n2}) using the $2 \times 2$ Hamiltonian instead of the $4 \times 4$ Hamiltonian, the resulting difference $\delta n$ has the same form as in Equation~(\ref{delta_n}). To compute $\Delta$ self-consistently, we use the same approach as described in section~\ref{SelfConsistentCalculations}A. The result for $\varepsilon_\text{z}=1.65$ is the grey line in Figures~3c and Figure~6 in the main text.\\

Since Equation~(\ref{2xHamiltonian}) considers only low-energy bands, the resulting $\Delta(D)$ for large $D$ values differs from the $\Delta$ values described in Ref.~\cite{Slizovskiy2021}.

\section{Bias-induced p-n-junctions and comparison of different samples by absolute resistance values}
Here we discuss how an applied bias voltage can significantly change the spatial profile of the band structure by creating a p-n-junction (compare cases A and C in Figure~4b of the main manuscript). As a consequence of the p-n-junction, the overall resistance of the device is dominated by a very small, line-shaped region somewhere between source and drain contacts. This justifies the comparison of different devices by absolute resistance values.\\

It is important to note that all equations in section~\ref{ChemicalPotential} of this Supplemental Material do not incorporate the effect of a bias voltage. Up to this point, it was implicitly assumed that the BLG is put to the same common ground potential to which all externally applied voltages (i.e. the gate voltages and the voltages applied to source and drain contacts) are referenced to. Instead, if the BLG is put to another potential because of an applied source-drain voltage, not the absolute gate voltages but rather the voltage differences between the gates and the BLG must be considered.\\

This gets important for case C in Figure~4b, where we apply the $V_\mathrm{sd}$ symmetrically over the gated BLG device, i.e. half the voltage is applied to the drain contact, and the other half is applied to the source contact with an inverted sign. Therefore, somewhere within the BLG device, there has to be one point that is on a virtual ground (i.e. a point that exhibits no voltage difference to the common ground potential). This point is marked with an orange triangle in Figure~4b. Only at this point, all previous equations hold and the resulting band gap energy and the position of $\mu$ are identical to case A, where the whole BLG channel is put to the common ground potential.\\

Without an applied $V_\mathrm{sd}$ (case A) the number of holes induced by the top gate and the number of electrons induced by the back gate exactly compensate each other and, therefore, no free charge carriers are present. Instead, in case C the voltage at the point marked with a blue triangle in Figure~4b will be $-V_\mathrm{sd}/2$ (we assume that the onset of conductance has been just reached and, therefore, the current is small enough that any current-induced voltage drop in the source and drain leads can be neglected). This voltage effectively reduces the voltage difference to the top gate by the same amount as it increases the voltage difference to the back gate. As a consequence, there will be a surplus of induced electrons, pushing the chemical potential into the conduction band. The exact opposite case occurs on the other side (red triangle in Figure~4b), where the voltage of the BLG channel is pushed towards the back gate voltage and, hence, the top gate now induces a surplus of holes.\\

Overall, the symmetrically applied $V_\mathrm{sd}$ creates a p-n-junction within the BLG channel, over which the vast majority of the $V_\mathrm{sd}$ drops as long as the current is sufficiently small. This creation of a p-n-junction explains the fact that an absolute resistance value is a good quantity for a comparison of different device technologies and not the resistivity that is normalized to the different device geometries (see Figs.~2 and~3 in the main manuscript). As the line-shaped p-n-junction dominates the overall transport, the total length of the transport channel is not important at all. The only variable is the width of the transport channel and, accordingly, the length of the p-n-junction from one edge of the transport channel to the other. And this width is in the same order of magnitude for the vast majority of devices presented in literature.\\

\subsection{Bias-induced p-n-junction including spatially varying disorder potential}
\begin{figure}[!h]
\includegraphics[draft=false,keepaspectratio=true,clip,width=1\linewidth]{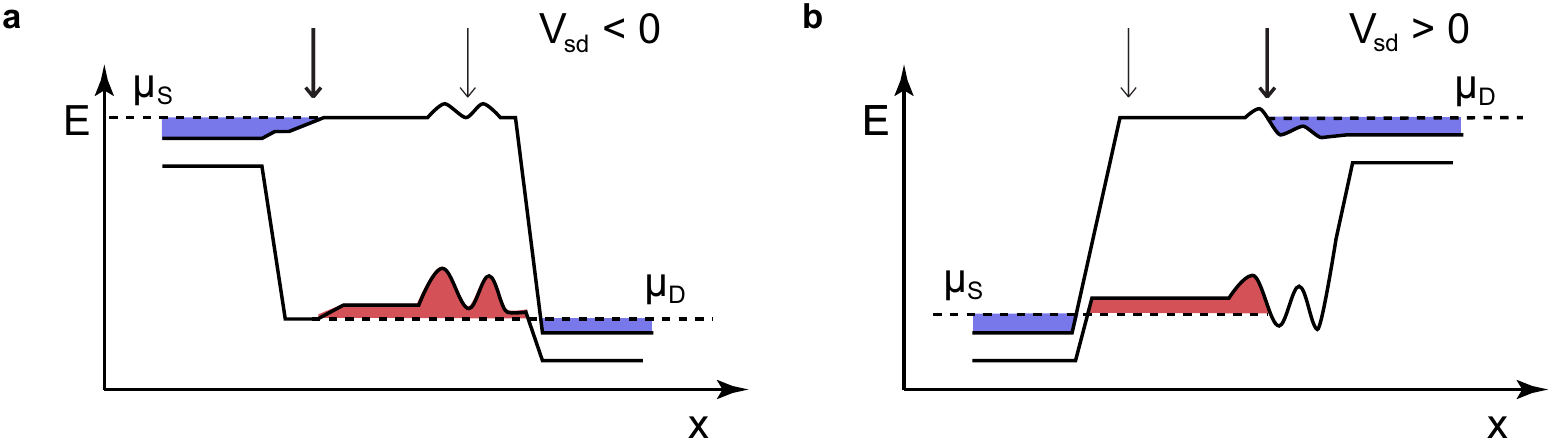}
\caption[Fig01]{Schematic representation of the transport regime C (compare to Figure~4 in the main text) for opposite polarities of the source-drain voltage V$_\text{sd}$ including potential disorder. The latter leads to the formation of the p-n junction at different positions along the gapped BLG channel (see bold arrows) depending on the sign of V$_\text{sd}$, resulting in possible asymmetric features in the bias spectroscopy.
}
\label{fig:r5}
\end{figure}

The disorder potential is expected to vary spatially over the whole bilayer graphene sample because of the nature of its origin (defects and contaminations at gate interfaces, etc.).
In Fig.~\ref{fig:r5} we show band edge diagramms in which we assume the presence of a  region with small disorder near the drain contact. The disorder is drawn in such a way that it is easier to induce holes than electrons in the disordered region by electrostatic gating. This affects where the p-n junction forms underneath the top gate for different polarities of the source-drain voltage $V_\text{sd}$. For $V_\text{sd}<0$, it is easy to flush the disordered region near the drain electrode with holes, making this region quite conductive, which in turn pushes the p-n junction towards the source electrode (denoted by the bold arrows in Fig.~\ref{fig:r5}). Instead, a positive bias voltage ($V_\text{sd}>0$) tries to push the region near the drain electrode into the electron regime, which is however hindered by the disorder, leading to the formation of the p-n junction closer to the drain contact. As a consequence, the bias spectroscopy will show some asymmetric features in the source-drain voltage $V_\text{sd}$ (see, e.g., Fig.~4a in the main text).

\bibliography{literature}